\documentclass{pasa}%

\usepackage{graphicx}
\usepackage{gensymb} 
\usepackage{xfrac}   
\usepackage[title]{appendix} 

\newcommand{\skacalcwebpage}{\url{http://sensitivity.skalow.link}}
\newcommand{\githubpage}{\url{https://github.com/marcinsokolowski/station\_beam}}
\newcommand{\red}[1] {#1}

\title[What is the SKA-Low Sensitivity]{What is the SKA-Low Sensitivity for Your Favourite Radio Source?}

\author[Sokolowski et al.]
{ M.~Sokolowski$^{1}$\thanks{marcin.sokolowski@curtin.edu.au}, 
  S.~J.~Tingay$^1$, 
  D.~B.~Davidson$^1$,
  R.~B.~Wayth$^{1,2}$,
  D.~Ung$^1$,
  J.~Broderick$^{1}$,
  B.~Juswardy$^{1}$,
  M.~Kovaleva$^{1}$,
  G.~Macario$^{3}$,
  G.~Pupillo$^{4}$,
  A.~Sutinjo$^{1}$
\affil{$^1$International Centre for Radio Astronomy Research, Curtin University, Bentley, WA 6102, Australia}%
\affil{$^2$ARC Centre of Excellence for All Sky Astrophysics in 3 Dimensions (ASTRO 3D), Australia}
\affil{$^3$Osservatorio Astrofisico di Arcetri, Istituto Nazionale di Astrofisica, Florence, Italy}
\affil{$^4$Istituto di Radioastronomia, Istituto Nazionale di Astrofisica, Via Gobetti, 101, Bologna, Italy}
}%

\jid{PASA}
\doi{10.1017/pas.\the\year.xxx}
\jyear{\the\year}

\usepackage{aas_macros}
\usepackage{hyperref} 
\hypersetup{colorlinks,citecolor=blue,linkcolor=blue,urlcolor=blue}

\hypersetup{draft}

\begin{document}

\begin{frontmatter}
\maketitle

\begin{abstract}
The Square Kilometre Array (SKA) will be the largest radio astronomy observatory ever built, providing unprecedented sensitivity over a very broad frequency band from 50\,MHz to 15.3\,GHz. The SKA's low frequency component (SKA-Low), which will observe in the $50 - 350$\,MHz band, will be built at the Murchison Radio-astronomy Observatory (MRO) in Western Australia. It will consist of 512 stations each composed of 256 dual-polarised antennas, and the sensitivity of an individual station is pivotal to the performance of the entire SKA-Low telescope. The answer to the question in the title is, it depends.  The sensitivity of a low frequency array, such as an SKA-Low station, depends strongly on the pointing direction of the digitally formed station beam and the local sidereal time (LST), and is different for the two orthogonal polarisations of the antennas. The accurate prediction of the SKA-Low sensitivity in an arbitrary direction in the sky is crucial for future observation planning. Here we present a sensitivity calculator for the SKA-Low radio telescope, using a database of pre-computed sensitivity values for two realisations of an SKA-Low station architecture.  One realisation uses the log-periodic antennas selected for SKA-Low.  The second uses a known benchmark, in the form of the bowtie dipoles of the Murchison Widefield Array. Prototype stations of both types were deployed at the MRO in 2019, and since then have been collecting commissioning and verification data. These data were used to measure the sensitivity of the stations at several frequencies and over at least 24\,h intervals, and were compared to the predictions described in this paper. The sensitivity values stored in the \textsc{SQLite} database were pre-computed for the X, Y and Stokes I polarisations in 10\,MHz frequency steps, \sfrac{1}{2} hour LST intervals, and 5\degree\, resolution in pointing directions. The database allows users to quickly and easily estimate the sensitivity of SKA-Low for arbitrary observing parameters (your favourite object) using interactive web-based or command line interfaces. 
The sensitivity can be calculated using publicly available web interface (\skacalcwebpage) or a command line \textsc{python} package (\url{https://github.com/marcinsokolowski/station\_beam}), which can also be used to calculate the sensitivity for arbitrary pointing directions, frequencies, and times without interpolations.
\end{abstract}

\end{frontmatter}


\section{INTRODUCTION}
\label{sec:intro}

The Square Kilometre Array \citep[SKA;][]{2009IEEEP..97.1482D}\footnote{https://www.skatelescope.org/} is a huge international endeavour to build the world's largest and most sensitive radio telescope, enabling observations at frequencies from 50\,MHz to 15.3\,GHz. The low frequency SKA (SKA-Low), covering the frequency range 50 - 350\,MHz, will be located in a radio-quiet zone, including the Murchison Radio-astronomy Observatory (MRO) in Western Australia. SKA-Low will address a very broad range of science cases, ranging from fundamental cosmological studies of the early Universe, the formation of the first stars and galaxies during the Cosmic Dawn, the study of the Epoch of Reionisation (when these first objects re-ionised the neutral hydrogen), the evolution of galaxies, cosmic magnetism, extraterrestrial life, astrophysical transients, general gravity, pulsars and black holes, high energy cosmic-rays and, very likely, completely new and unexpected discoveries \citep{Braun:2015B3}. All the SKA-Low science goals rely heavily on its unprecedented broad-band sensitivity, which will result from the enormous number of individual antennas (131,072) providing an effective area of the order of a square kilometre.

SKA-Low will consist of 512 stations, each composed of 256 dual-polarised antennas. Electrical signals from all individual antennas within each station are digitised and coherently added in a digital beamformer. The resulting data streams (one from each of the two antenna polarisations) of high time resolution station beam complex voltages are transported to the Central Processing Facility via optical fibres where they are cross-correlated between all stations. Therefore, the sensitivity of the digitally formed individual station beams is critical to the sensitivity of the entire SKA-Low telescope and, consequently, the realisation of its science goals. The design specifications for the SKA-Low \red{(Phase 1)} stations, including sensitivity specifications, are defined in \citep{SKA1L1req}.

Four full-scale prototype SKA-Low stations (composed of 256 dual-polarised antennas) have already been deployed at the MRO since 2016. Starting with the Engineering Development Array 1 \citep[EDA1;][]{2017PASA...34...34W}, which began its operations in the early 2016 to verify performance of the bowtie dipoles and analogue beamformers used in SKA-Low precursor the Murchison Widefield Array \citep[MWA;][]{2013PASA...30....7T}. Around the same time, the Aperture Array Verification System 1 \citep[AAVS1;][]{aavs1_paper}, consisting of 256 individually digitised SKALA2 antennas, \citep{8520395,2017MNRAS.469.2662D} was deployed at the MRO. 

Based on these experiences, two next generation prototype stations were deployed in 2019, the Engineering Development Array 2 \cite{eda2_paper} and the Aperture Array Verification System 2 \citep[AAVS2;][]{andre_spie,giulia_et_al}, were deployed at the MRO. They use the same signal chain technologies and the antenna layout as the AAVS1, but the antenna designs in both stations are different. The EDA2 station, as for its predecessor (EDA1), consists of 256 MWA bowtie dipoles, whilst the AAVS2 station is composed of SKALA4.1AL antennas \citep{9107113}. These two stations were built in order to verify the digital technology, the sensitivity of the SKA-Low stations using different antenna designs and to compare their performance against each other and the SKA-Low requirements \citep[defined in][]{SKA1L1req}.

The most recent theoretical analysis of the SKA-Low performance expectations, in terms of sensitivity, is provided in \cite{2019arXiv191212699B}.  In summary, the expectations from \cite{2019arXiv191212699B} are based on an early version of the SKALA antenna and SKA-Low station configuration, with sensitivity values obtained from averages over all solid angles within 45$^{\circ}$ from zenith and referenced to an assumed Galactic foreground contribution, with a 408 MHz sky temperature of 20 K scaled in frequency according to $\propto (408/\nu_{MHz})^{2.75}$. This corresponds to a value between the 10$^{\rm th}$ and 50$^{\rm th}$ percentile of the all-sky temperature distribution and would apply to directions well away from the Galactic plane.

\red{The approximations used by \citet{2019arXiv191212699B}, that is uniform sky temperature and sensitivity averaged over elevations $\ge$45\degree,} while useful to provide an order-of-magnitude average sensitivity, are far from what is required in order to assess detailed sensitivity predictions for particular science targets or programs of observation.  The nature of low frequency aperture arrays means that the achievable sensitivity is a function of position on the sky, due to a strongly variable sky temperature distribution.  The response of a complex, broadband antenna to the sky also changes strongly as a function of frequency and pointing direction.  Furthermore, the two orthogonal antenna polarisations also respond to the same sky differently, meaning that the total Stokes I sensitivity can be a complex combination of the X and Y polarisation measurements \citep{2021A&A...646A.143S}.

We seek to significantly advance the understanding of the SKA-Low sensitivity, for particular observation scenarios, that will assist astronomers interested in planning and assessing SKA-Low observation programs.  To this end, this paper describes software tools enabling astronomers and engineers to very quickly calculate the expected sensitivity of an SKA-Low station in X, Y, and Stokes I polarisations at an arbitrary observing frequency (within the $50 - 350$\,MHz band), time, and pointing direction. The sensitivity values are pre-computed and have been saved to an \textsc{SQLite} database\footnote{https://www.sqlite.org/index.html} for both station designs (the EDA2 and AAVS2) at $5\degree$ spatial resolution, \sfrac{1}{2} hour intervals in local sidereal time (LST), and 10\,MHz frequency steps. The sensitivity can be retrieved from the database using a \textsc{DJANGO} web-service application available at \skacalcwebpage \, or command line \textsc{python} script \texttt{eda\_sensitivity.py} available at \githubpage. The \textsc{python} script \texttt{eda\_sensitivity.py} can be used to calculate sensitivity at an arbitrary frequency, time, and pointing direction (other than pre-computed values stored in the database), which takes approximately 8\,seconds for a single time, frequency, and pointing direction.

This paper is organised as follows. In Section~\ref{sec:sensitivity_simulations} we present the method and software used to efficiently calculate a large number of sensitivity values for the SKA-Low prototype stations AAVS2 and EDA2. In Section~\ref{sec:sensitivity_database} we describe the sensitivity database and how it was populated with sensitivity values pre-computed for the two stations. In Section~\ref{sec:sensitivity_calculator} we present the web-based and command line sensitivity calculators enabling access to the pre-calculated or interpolated sensitivity values, and the calculation of sensitivity for arbitrary observing parameters. In Section~\ref{sec:eda2_aavs2_comparison} we summarise the comparisons between the sensitivities predicted by the presented software and measurements described in other publications. Finally, in Section~\ref{sec:discussion} we make summary remarks and outline future plans. 

\section{SKA-Low station sensitivity simulations}
\label{sec:sensitivity_simulations}

The sensitivity of a radio telescope in one of the instrumental linear polarisations (X or Y) at frequency $\nu$ and a given local sidereal time (LST) can be calculated as $AoT(\nu,\text{LST}) = A_e(\nu)/T_{sys}(\nu,\text{LST})$, where $A_e(\nu)$ is the LST-independent effective area of an aperture array (or a dish) and $T_{sys}(\nu,\text{LST})$ is the system temperature representing the system noise -- a sum of the noise contributions from the sky and receiver. In principle $A_e(\nu)$ should be time independent. We note that all these quantities should also have polarisation index p (X or Y), which was dropped for brevity.

\subsection{Calculation of system temperature}
\label{sec:calc_system_temperature}

\begin{figure}   	
   	\includegraphics[width=\columnwidth]{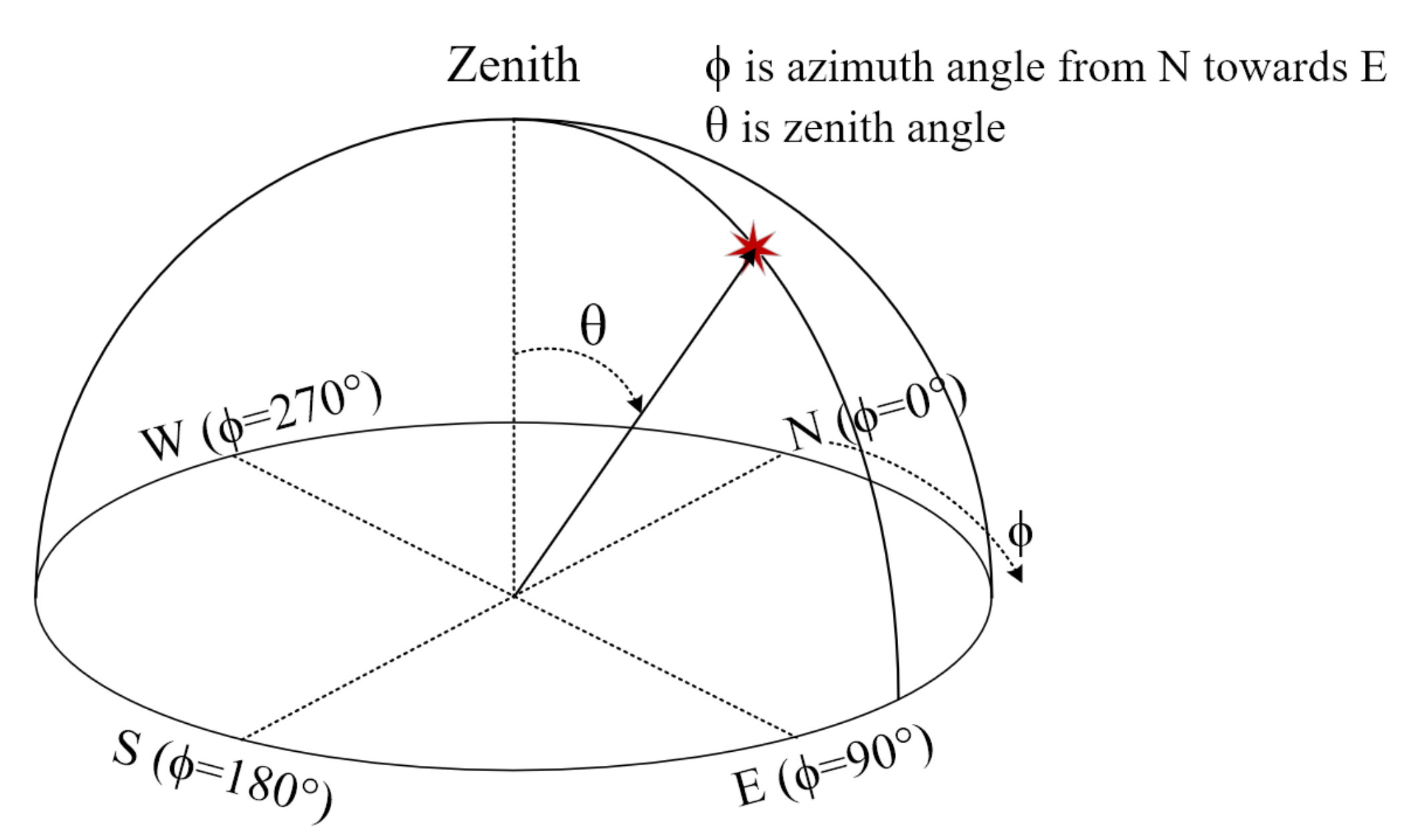}
    \caption{\red{The horizontal coordinates system used in the paper with the definitions of the angles $\theta$ (zenith angle) and $\phi$ (azimuth starting from the North at $\phi=$0\degree \,and increasing towards the East at $\phi=$90\degree).}}
    \label{fig_coordinate_system}
\end{figure}

\begin{figure}
   	\includegraphics[width=\columnwidth]{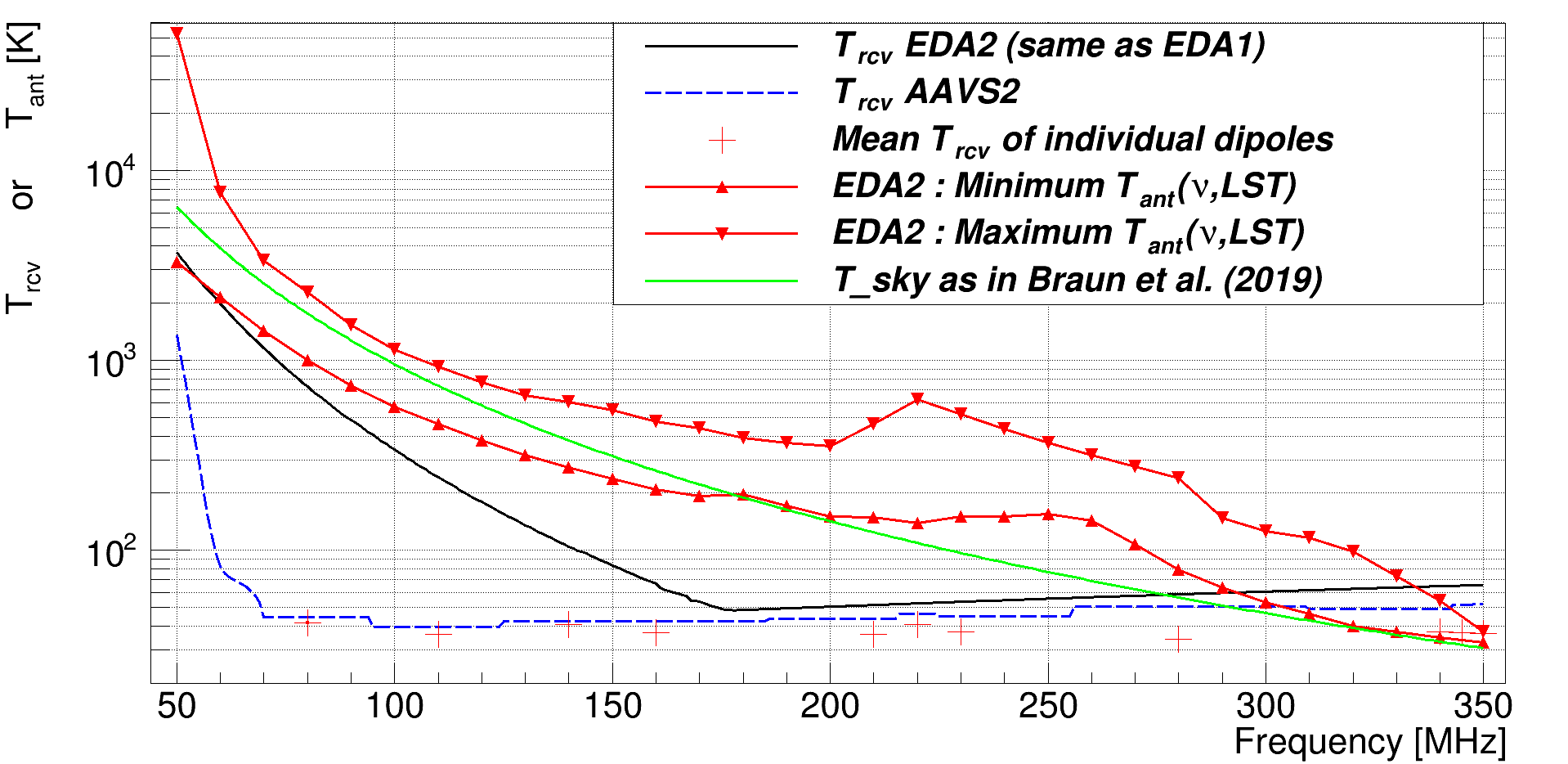}
    \caption{The receiver temperature of the AAVS2 at zenith (dashed blue line) and EDA2 (solid black line) stations used in the simulations presented in this paper. The red crosses were calculated as the mean over receiver temperatures of all individual antennas in the AAVS2 station as simulated in FEKO. They are very similar to the values used for the AAVS2 (dashed blue line). For comparison, the red solid curves with triangles pointing up and down show, are respectively, the minimum and maximum $T_{ant}(\nu,\text{LST})$ calculated over all LSTs for the EDA2 station (the curve for AAVS2 is virtually the same and was not plotted for the clarity of the image). \red{The green curve is the $T_{sky}$ used by \citet{2019arXiv191212699B}.}}
    \label{fig_receiver_temperature}
\end{figure}

\begin{figure}
   	\includegraphics[width=\columnwidth]{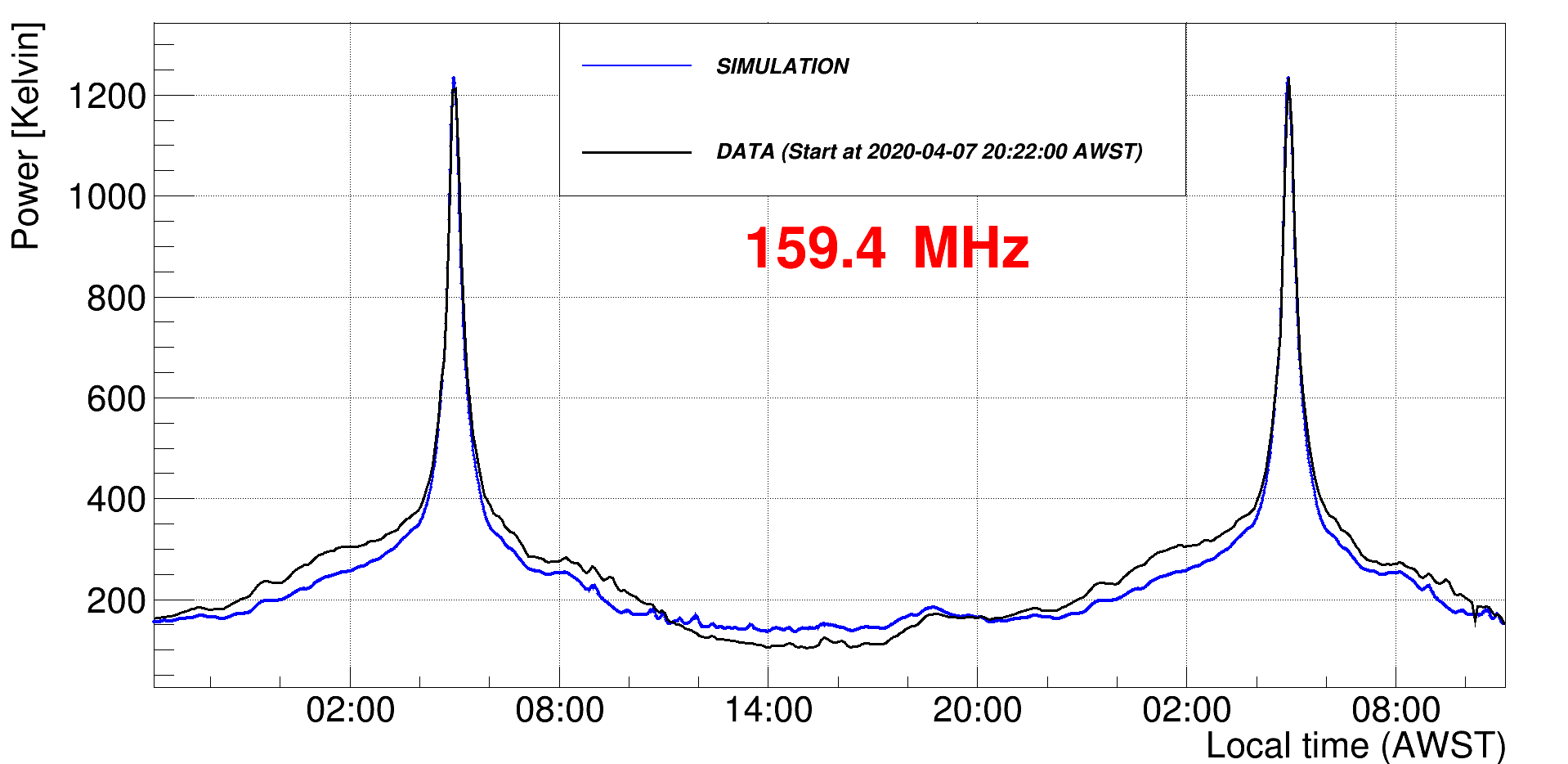}
    \caption{Total power recorded with the AAVS2 station beam pointed at the zenith (drift scan observation) as a function of local time at a frequency of 159.375\,MHz. The uncalibrated data were normalised to the simulation (in Kelvin units) at the peak value. The simulation matches the data points very well (within 10-20\% for most of the observation duration). The approximate six-fold change in antenna temperature at different LST times causes correspondingly large changes in the station sensitivity (see examples in Figures~\ref{fig_aavs2_sens_vs_time_azim0_za0} and \ref{fig_eda2_sens_vs_time_azim0_za0}).}
    \label{fig_aavs2_drift_scan_ch204}
\end{figure}

System temperature, $T_{sys}(\nu,\text{LST})$, is a sum of receiver temperature ($T_{rcv}(\nu)$) and antenna temperature ($T_{ant}(\nu,\text{LST})$), i.e. $T_{sys}(\nu,\text{LST}) = T_{ant}(\nu,\text{LST}) + T_{rcv}(\nu)$. The receiver temperature, $T_{rcv}(\nu)$, shown in Figure~\ref{fig_receiver_temperature}, is time independent. Therefore, it is straightforward to note that the sensitivity of the low frequency array depends not only on its collecting area ($A_e(\nu)$), but also on the system temperature ($T_{sys}(\nu,\text{LST})$), which is strongly LST-dependent due to its sky noise component, $T_{ant}(\nu,\text{LST})$. The strong LST-dependence of system temperature can be seen in Figure~\ref{fig_aavs2_drift_scan_ch204} showing the total power ($P(\nu,\text{LST})$), recorded by the AAVS2 station pointed at zenith during a 48\,h observation (drift scan observation). 

This total power is directly proportional to the system temperature, $T_{sys}(\nu,\text{LST})$:

\begin{equation}
P(\nu,\text{LST}) = g(\nu,\text{LST}) \left[ T_{ant}(\nu,\text{LST}) + T_{rcv}(\nu) \right],
\label{eq:power_tsys}
\end{equation}

where $g(\nu,\text{LST})$ is the frequency dependent gain of the system, which can also change in time due to variations in gains of various components. In our simulations, the antenna temperature, $T_{ant}(\nu,\text{LST})$, was calculated as an integral of sky temperature ($T_{sky}(\nu,\theta,\phi, \text{LST})$) weighted by the beam pattern of the station beam:

\begin{equation}
T_{ant}(\nu,\text{LST}) = \frac{\int_{4\pi} B_{st}(\nu,\theta,\phi) T_{sky}(\nu,\theta,\phi,\text{LST}) d\Omega}{\int_{4\pi} B_{st}(\nu,\theta,\phi) d\Omega}, 
\label{eq:sky_integration}
\end{equation}
where $B_{st}(\nu,\theta,\phi)$ is the beam pattern of an SKA-Low station (EDA2 or AAVS2), $T_{sky}(\nu,\theta,\phi,\text{LST})$ is the sky brightness temperature from the sky model at frequency $\nu$ and pointing direction $(\theta,\phi)$ \red{in horizontal coordinates}, which implies LST dependence as the sky above the horizon changes as the Earth rotates. \red{The angles $\theta$ and $\phi$ are respectively zenith angle and azimuth starting from the North and increasing towards the East (Fig.~\ref{fig_coordinate_system})}. The sky model used in our calculations is based on \citet{1982A&AS...47....1H} at 408\,MHz (the ``Haslam map'') scaled to lower frequencies using a spectral index of $-2.55$ \citep{spectral_indexASU2019}. 

An example comparison of the measured total power of the station beam in the X polarisation pointed to the zenith (i.e. drift scan observation) and our simulation is shown in Figure~\ref{fig_aavs2_drift_scan_ch204}. The data and simulation are normalised at the peak value. Figure~\ref{fig_aavs2_drift_scan_ch204} shows that the antenna temperature ($T_{ant}(\nu,\text{LST})$) varies between approximately 200\,K and 1200\,K (by a factor of six). Moreover, the simulation matches the data very well with residuals within $\pm$10-15\% over almost 48\,hours.

As shown in Figure~\ref{fig_receiver_temperature}, in the case of SKA-Low prototype stations AAVS2 and EDA2, at the time of Galactic Centre transit (LST$\approx17.8$\,h), the system temperature at frequencies below 330\,MHz is dominated by the antenna temperature (even by an order of magnitude at frequencies $\le$250\,MHz). However, even when the Galactic Centre is at its lowest point below the horizon (``cold sky'' is above the horizon) the antenna temperature still exceeds the receiver temperature below 300\,MHz for both AAVS2 and EDA2 stations, but typically by less than an order of magnitude (Fig.~\ref{fig_receiver_temperature}). Therefore, through the dependence on $T_{ant}(\nu,\text{LST})$, the station sensitivity strongly depends on the pointing direction (``cold'' vs. ``hot'' parts of the sky) and the LST of the observations. \red{For comparison, the green curve in Figure~\ref{fig_receiver_temperature} is the $T_{sky}$ as calculated by \citet{2019arXiv191212699B}, which is between the ``cold'' and ``hot'' sky up to $\sim$180\,MHz and slightly below the temperature corresponding to ``cold'' sky at higher frequencies. This shows that $T_{sky}$ used by \citet{2019arXiv191212699B} can often be a factor of a few (in extremes even an order of magnitude) under- or over-estimated, resulting in a similar inaccuracy in the predicted sensitivity.}

The receiver temperature of the EDA2 station, which is the same as for the EDA1, was obtained from astronomical measurements \citep{2017PASA...34...34W} and confirmed with laboratory measurements by \citet{9040892}; both sets of measurements are shown in Figure~\ref{fig_receiver_temperature}. However, for the AAVS2 station, a model of zenith dependent receiver temperature was derived based only on laboratory measurements using the same technique as \citep{9040892}, and may be subject to modifications once it is also measured using astronomical measurements. All the above values ($T_{ant}(\nu,\text{LST})$ and $T_{rcv}(\nu)$) are stored in the database. Therefore, if the receiver temperature changes (for instance due to improved measurement or modifications in the receiver) it is straightforward to update the sensitivity values in the database by dividing effective area by a sum of $T_{ant}$ and the updated $T_{rcv}$, without the need for performing the simulations again.

\subsection{Calculation of station effective area}
\label{sec:calc_station_eff_area}

%
The effective area, $A_e(\nu,\theta_p,\phi_p)$, of the station at pointing direction $(\theta_p,\phi_p)$ can be calculated according to the standard equation:

\begin{equation}
A_e(\nu,\theta_p,\phi_p) = \frac{G(\nu,\theta_p,\phi_p)c^2}{4\pi\nu^2}, 
\label{eq_aeff}
\end{equation}

where $c$ is the speed of light and $G(\nu,\theta_p,\phi_p)$ is the station gain calculated as:

\begin{equation}
G(\nu,\theta_p,\phi_p) = \frac{4\pi B_{max}(\theta_p,\phi_p) \eta}{\int_{2\pi} B_{st}(\theta,\phi) \delta \Omega},
\label{eq_aeff}
\end{equation}

where $B_{max}(\theta_p,\phi_p)$ is the maximum value of the station beam pattern $B_{st}(\nu,\theta,\phi)$ over the entire sky when the station beam is pointed in the direction $(\theta_p,\phi_p)$ and radiation efficiency, $\eta$, was assumed to be $\eta=$1 in our calculations. Further considerations assume that the station beam is pointed in the direction $(\theta_p,\phi_p)$ but these indexes have been dropped for brevity. Typically, $B_{st}(\theta,\phi)$ has a maximum value at zenith ($\theta=0$\degree). However, we note, that this is not the case for the EDA2 (using MWA bowtie dipoles) at frequencies above 200\,MHz where the maximum response of a single dipole can be away from zenith. Therefore, the sensitivity at zenith can be lower than the maximum sensitivity at a given frequency.

The array gain can be accurately estimated with the electromagnetic simulations of the station beam at a given frequency and pointing direction using the Method of Moments implemented in electromagnetic simulation packages such as FEKO\footnote{https://altairhyperworks.com/product/FEKO} or Gallileo\footnote{https://www.idscorporation.com/pf/galileo-suite/}. This is a very accurate method, which takes into account the mutual coupling of the antennas within the stations. However, its main limitation, is that it is very computationally demanding and takes a relatively long time to simulate the beam at a single pointing direction and frequency. For frequencies above 80\,MHz it takes between 1 and 4 days to process a single frequency channel and polarisation, whereas below 80\,MHz it can take even between 7 and 14 days to simulate the station beam of AAVS2 at one pointing direction. Consequently, full electromagnetic simulations of both stations at 10\,MHz resolution were not available at the time of EDA2 and AAVS2 sensitivity measurements and verifications. 

\begin{figure*}
   	\includegraphics[width=\textwidth]{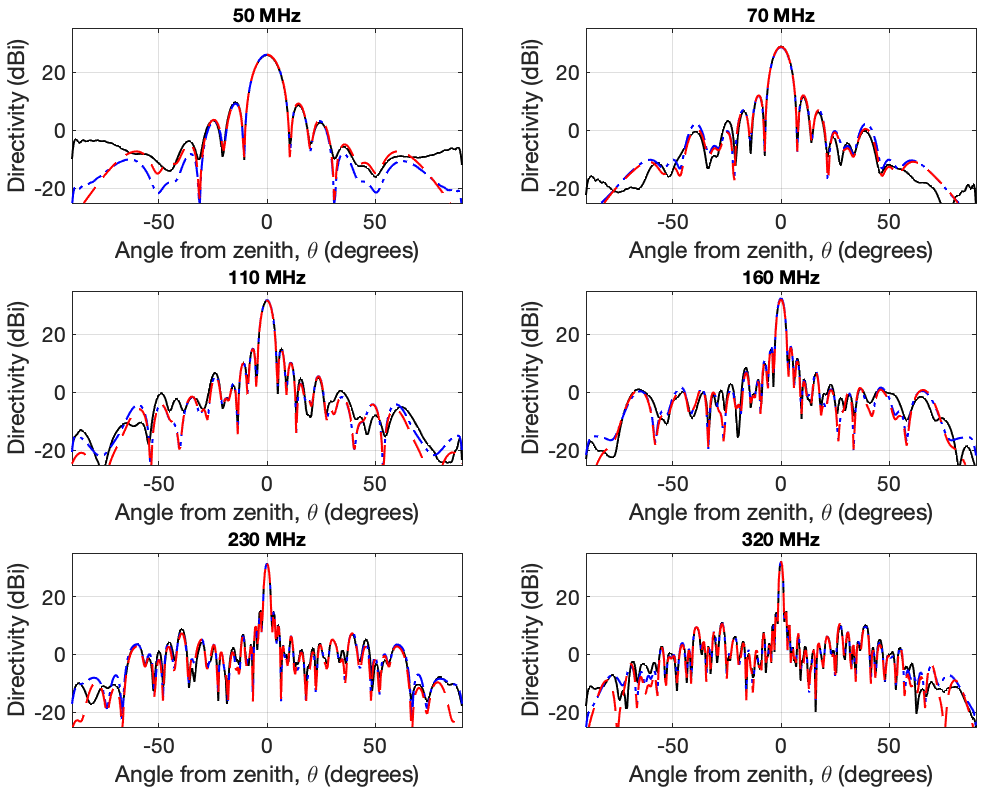} %
    \caption{A comparison of the AAVS2 station beam formed from: the array factor and the average element pattern (blue dash-dot curve); the array factor and the isolated element pattern (red  dashed curve); and a rigorous EEP formulation (black solid curve). This is for the X-polarisation, E-plane (i.e. in the plane of the dipoles).}
    \label{fig_EplanePat}
\end{figure*}

Therefore, a simplified array factor (AF) method has been used, which does not take into account mutual coupling effects. This method, which is the standard approach in antenna texts for array analysis \citep{BalanisAntenna3rd}, will be referred to as the array factor, or AF method. A more detailed description can be found in Section III of \citet{9410962} and will be briefly summarised here. In general, the array factor (eq. 4 in \citet{9410962}) is multiplied by the beam response of a single antenna element (eq.~5 in \citet{9410962}), which in our case is a beam pattern of an individual antenna within a station. The beam pattern for a single isolated element (ISO) over an infinite ground screen was simulated with the FEKO electromagnetic simulation software. This simulation was performed in 1\,MHz steps, enabling us to calculate sensitivity at many more frequency channels than using full electromagnetic simulations. Alternatively, we could have used average embedded element (AEE) pattern (average beam pattern of all dipoles within a station). However, the calculation of the AEE patterns requires the calculation of 256 embedded elements patterns (EEPs) which, as mentioned earlier, were only available at a few selected frequencies due to computational complexity. The difference between using AEE and ISO patterns was tested at these frequencies and the differences in the resulting sensitivities are within 15\%. Hence, our method is a combination of mathematically calculated array factor and FEKO simulations of a single antenna element. Results shown in Fig.~\ref{fig_EplanePat} indicate that the station main beam can be reliably formed using approximate  element patterns --- both average and isolated element patterns are shown --- at least in terms of the main beam and the inner sidelobes. The reason is that the array factor (geometric phase delays) dominates the element patterns by one to two orders of magnitude in the main beam.  However, this is no longer true as one moves further into the sidelobes. More details of the electromagnetic simulation of AAVS2 may be found in \citet{davids_paper}.
We note that once full electromagnetic simulations are feasible and available, the sensitivities in the database can be updated with even more precise values.
%
%


\subsection{Calculation of the system equivalent flux density}
\label{sec:calc_sefd}

The simulation software is based on simulations originally developed in 2016 to verify the sensitivity of the EDA1. The original code used an analytical model of the MWA dipole and this code has been modified to enable use of an arbitrary beam pattern for an individual element, which has to be provided in the FITS file format. The individual dipole patterns were simulated as a single SKALA4.1AL antenna or an MWA bowtie dipole over an infinite ground-screen.

System equivalent flux density (SEFD) is the flux density incident on the antenna (or aperture array) resulting in delivered power equal to the system noise ($A_e(\nu) \cdot $SEFD = $2kT_{sys}(\nu,\text{LST})$), where k is the Boltzmann constant. Hence, the SEFD = 2$kT_{sys}(\nu,\text{LST}) / A_e(\nu)$ can be calculated for X and Y polarisations by calculating $T_{sys}(\nu,\text{LST})$ and $A_e(\nu)$ for a specific instrumental polarisation (X or Y). 
As shown by \citet{2021A&A...646A.143S} this approach is only valid for unpolarised sources, while the calculation of SEFD for polarised sources is not straightforward. We used the following equation to calculate SEFD in Stokes I polarisation:

\begin{equation}
\text{SEFD}_{I} =  \frac{1}{2}\sqrt{\text{SEFD}_{XX}^2+\text{SEFD}_{YY}^2},
\label{eq:sefd_i}
\end{equation}

where $\text{SEFD}_{XX}, \text{SEFD}_{YY}$, and $\text{SEFD}_{I}$ are SEFDs in X, Y, and Stokes I polarisations, respectively. However, as discussed in \citet{2021A&A...646A.143S}, this equation is only valid in the cardinal planes (azimuth values 0\degree, 90\degree, 180\degree\, and 270\degree) and leads to errors for the off-zenith pointing directions (up to 40\% below an elevation of 30\degree\, in the MWA case). Nevertheless, we used Equation~\ref{eq:sefd_i} in the current version of the database.  Once more accurate equations (valid over the entire hemisphere) for the relationship between the X and Y polarisations and Stokes I are derived, the sensitivity in Stokes I will be updated accordingly.

At this stage, the software provides SEFD values, and it does not calculate standard deviation of the noise ($\sigma$) expected in the images of the sky. This functionality may be added in the future versions, but the expected $\sigma$ of the noise can be calculated using equation 9.35 in \citet{2009tra..book.....W}, which we repeat below:

\begin{equation}
\sigma = \frac{M \cdot SEFD}{\sqrt{N_{st}(N_{st}-1)\Delta t \Delta \nu}}, 
\label{eq:rms}
\end{equation}

where $N_{st}$ is the number of SKA-Low stations (expected to be 512), $\Delta t$ is the integration time in seconds, $\Delta \nu$ is the bandwidth in Hz and $M$ is a factor accounting for additional noise due to analogue to digital conversions (can be assumed to be equal to one).


\section{SKA-Low station sensitivity database}
\label{sec:sensitivity_database}

Our \textsc{python} package enables the calculation of sensitivity at arbitrary pointing directions, observing times, and frequency resolution same as a FEKO simulation of the individual antenna element above an infinite ground-screen (1\,MHz). In the present version, sensitivity values were pre-computed in 10\,MHz steps at frequencies from 50 to 350\,MHz, in \sfrac{1}{2} hour LST intervals, and 5\degree\, pointing direction resolution and saved to a \textsc{SQLite} database. The values tabulated in the database can be interpolated to calculate sensitivity at arbitrary time, frequency, and pointing direction and the resulting errors are within 10\%. Having the database of pre-computed values helps to avoid the re-calculation of sensitivity multiple times for the same (or nearly same) sets of observing parameters. Moreover, it enabled development of the web interface which returns the required sensitivity estimates within seconds rather than minutes (approximately 280\,s for 35 frequency channels at a single LST time and pointing direction) or hours (at multiple LSTs) required to calculate sensitivity using the underlying \textsc{python} simulation package. 

\subsection{Structure of the database}
\label{sec:database_structure}

\begin{figure}
   	\includegraphics[width=\columnwidth]{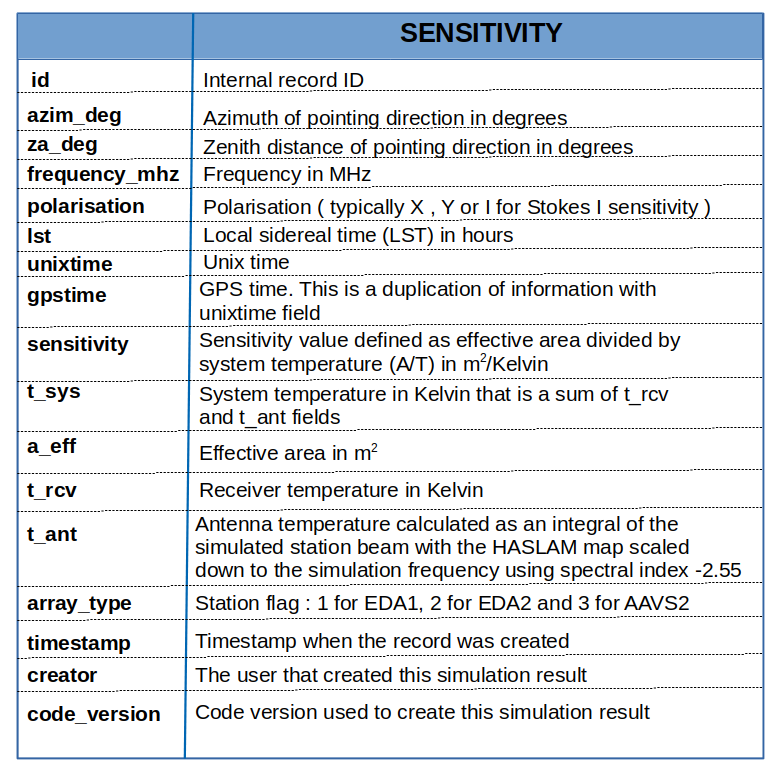} %
    \caption{The definition of the Sensitivity table in the database.}
    \label{fig_sensitivity_db}
\end{figure}

The database consists of a single table and its structure is shown in Figure~\ref{fig_sensitivity_db}. 
The sensitivity values for both stations were calculated on the cloud computing system ``Nimbus'' hosted by the Pawsey Supercomputing Centre (PSC). The calculated values were saved to \textsc{SQL} files and uploaded to the \textsc{SQLite} database (\textsc{text} files were also archived for the future reference). The size of the database file for a single station is approximately 670\,MB and contains 4,354,630 records resulting from 48 half hour LST steps $\times$ 35 frequencies $\times$ 18 elevations $\times$ 72 azimuths $\times$ 2 polarisations + 70 (35 frequency points in 2 polarisations at zenith where the sensitivity was calculated at a single azimuth $=$ 0\degree). 

\section{Sensitivity calculator}
\label{sec:sensitivity_calculator}

Software tools providing access to the sensitivity database have also been implemented in \textsc{python} and consist of a command line script and web interface, which will be described in this section. They both use the pre-computed sensitivity values stored in \textsc{SQLite} database files (separate file for each station). Moreover, the package also contains the original sensitivity simulation \textsc{python} script \texttt{eda\_sensitivity.py}\footnote{originally developed for the EDA1 simulations hence the name}, which can be used to calculate SKA-Low station sensitivity at arbitrary frequency (at integer MHz values), time, and pointing direction. The manual for this script is included in the \textsc{github} repository.

\subsection{Command line tools}
\label{subsec:command_line}

\begin{figure*}
   	\includegraphics[width=\columnwidth]{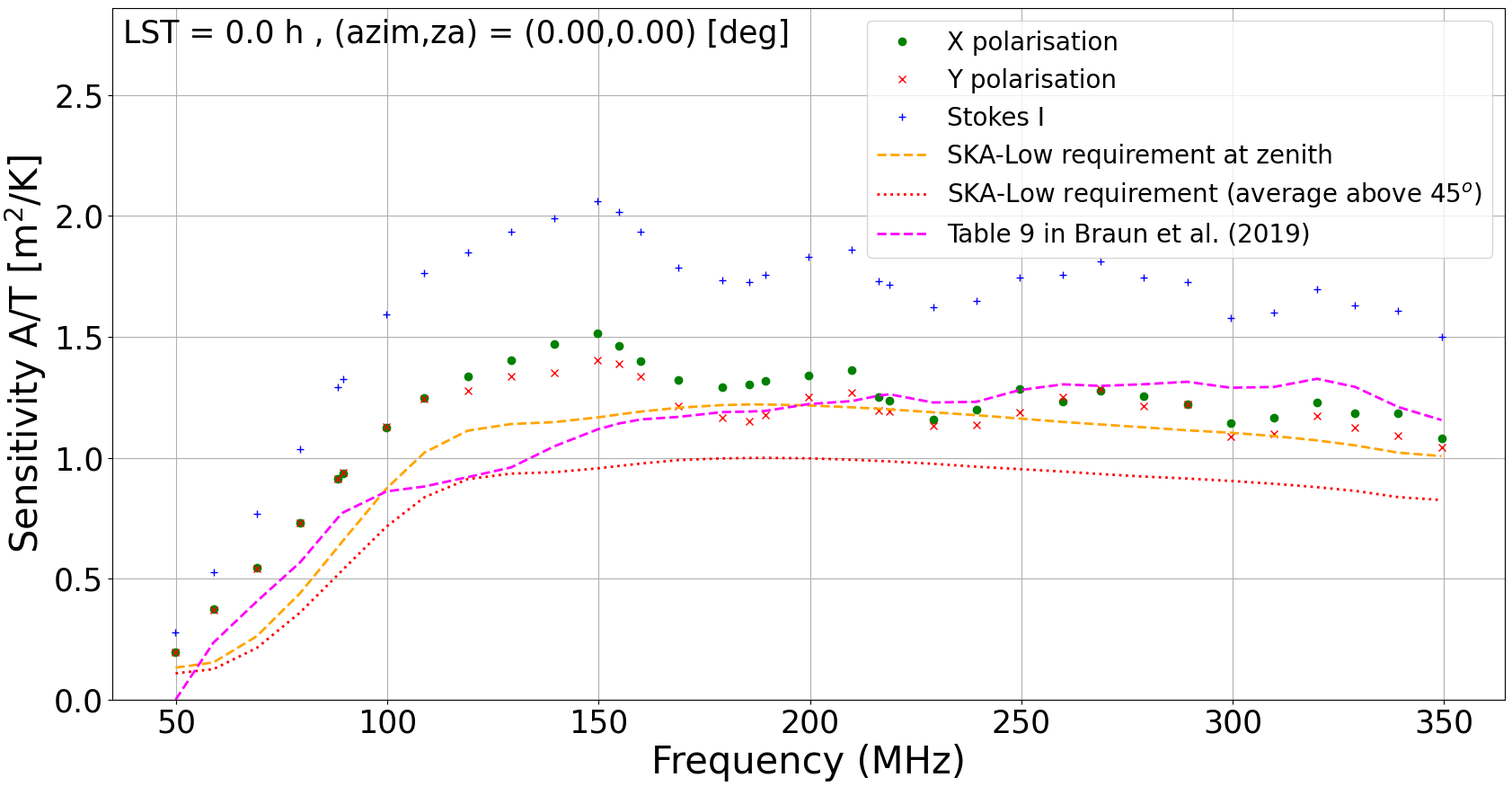} %
   	\includegraphics[width=\columnwidth]{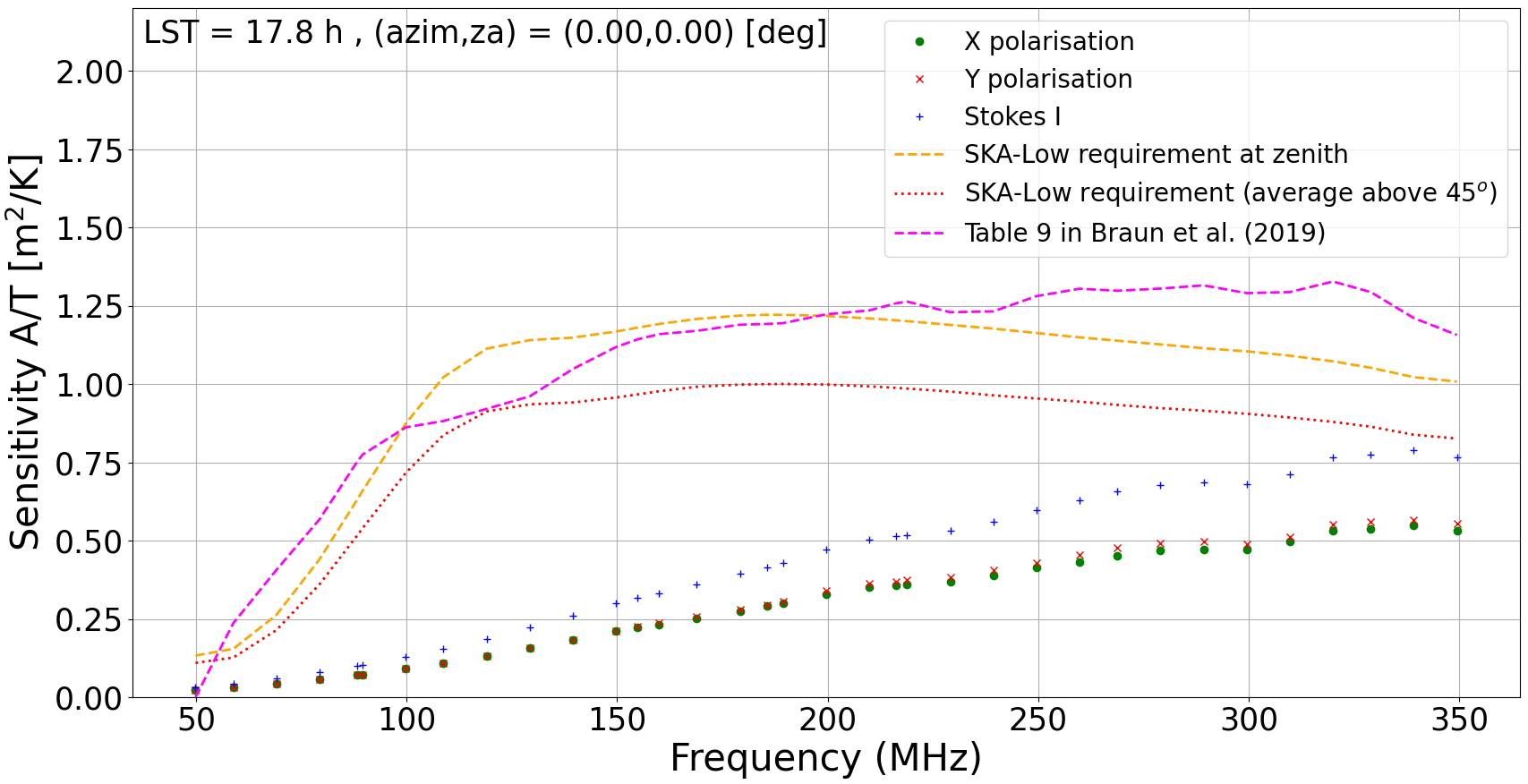} %
    \caption{The AAVS2 sensitivity as a function of frequency at the zenith pointing. Left image : the ``cold sky'' is transiting
    at LST = 0\,h. Right image : Galactic transit at LST = 17.8 \,h. \red{The data points are results of this work, dashed orange and dotted red curves are SKA-Low specifications at zenith and averaged over elevations $\ge$45\degree\,, respectively. The magenta dotted curve is the sensitivity from Table 9 in \citet{2019arXiv191212699B} (also averaged over elevations $\ge$45\degree).}}
    \label{fig_aavs2_sens_vs_freq_azim0_za0}
\end{figure*}

\begin{figure*}
   	\includegraphics[width=\columnwidth]{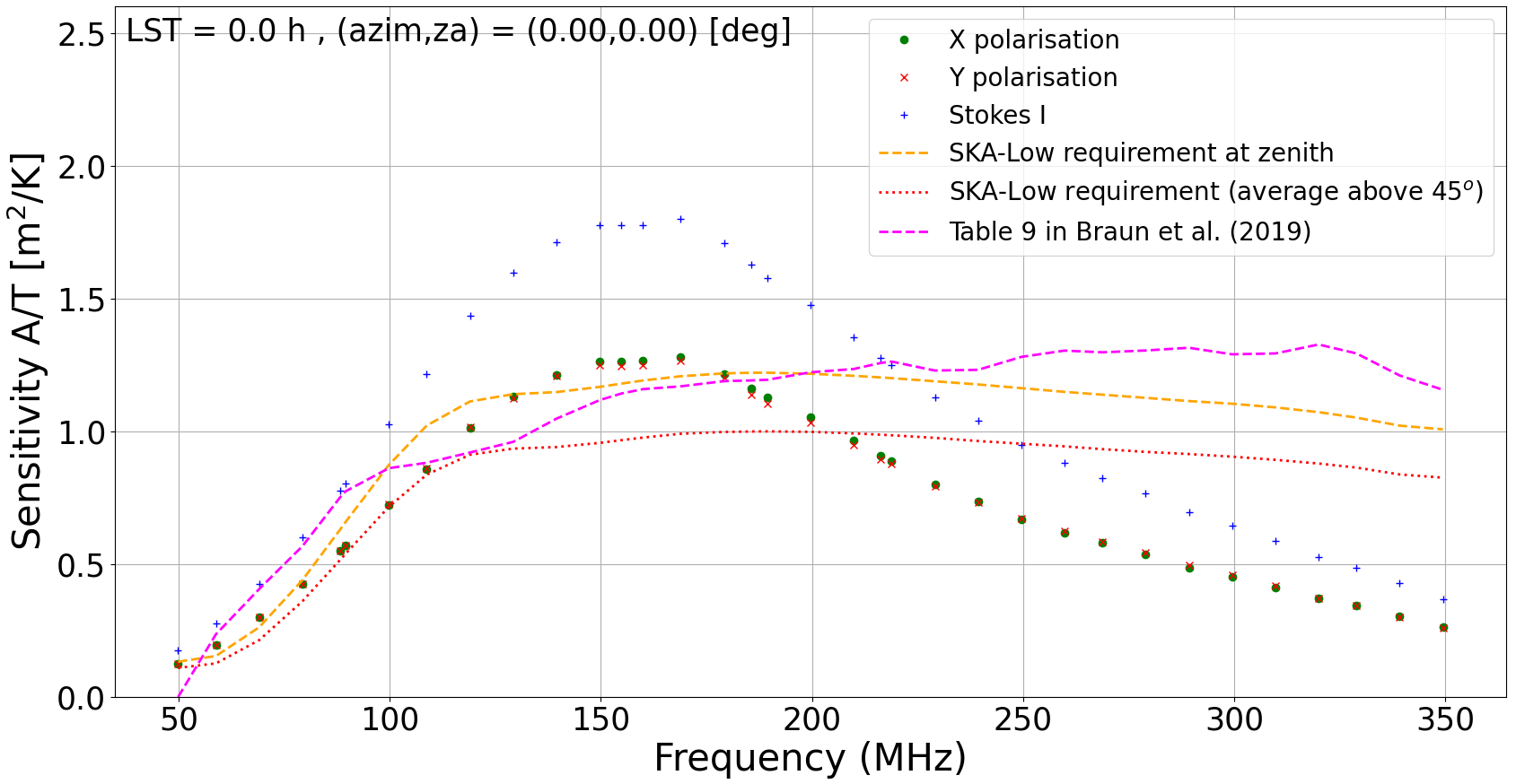} %
   	\includegraphics[width=\columnwidth]{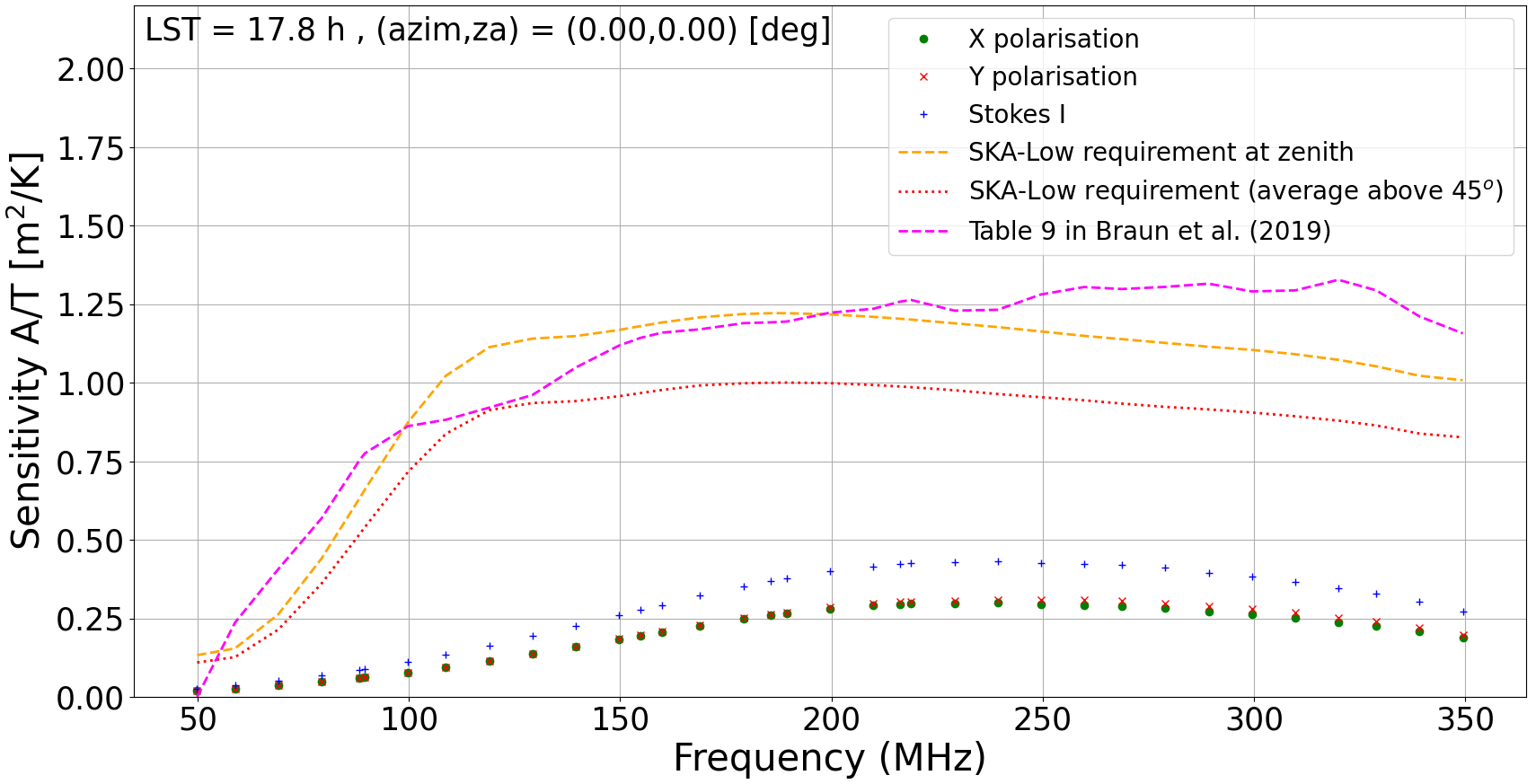} %
    \caption{The EDA2 sensitivity as a function of frequency at the zenith pointing. Left image : the ``cold sky'' is transiting at LST = 0\,h. Right image : Galactic transit at LST = 17.8 \,h. \red{The data points are results of this work, dashed orange and dotted red curves are SKA-Low specifications at zenith and averaged over elevations $\ge$45\degree\,, respectively. The magenta dotted curve is the sensitivity from Table 9 in \citet{2019arXiv191212699B} (also averaged over elevations $\ge$45\degree).}}
    \label{fig_eda2_sens_vs_freq_azim0_za0}
\end{figure*}

\begin{figure*}
   	\includegraphics[width=0.35\textwidth]{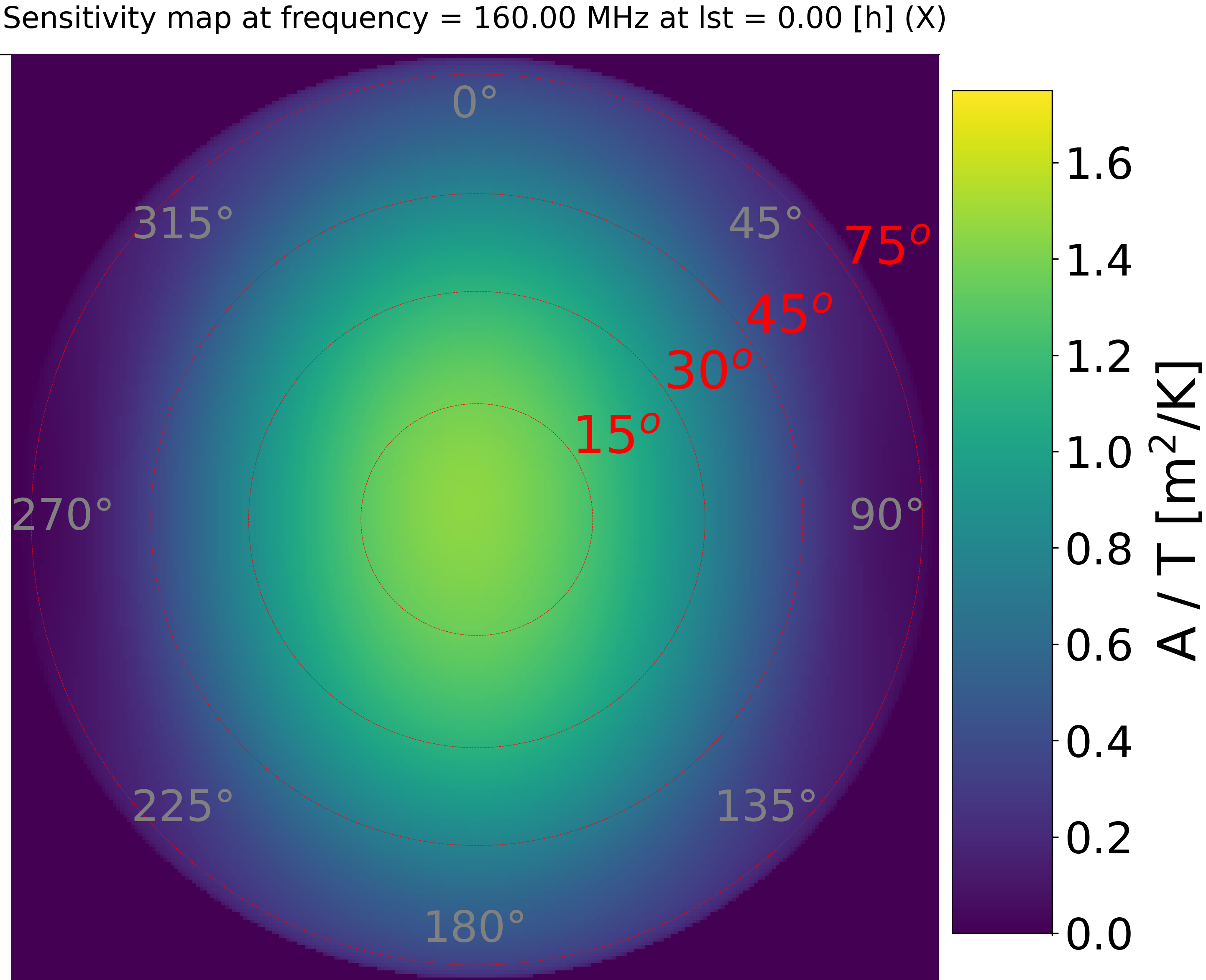} %
   	\includegraphics[width=0.35\textwidth]{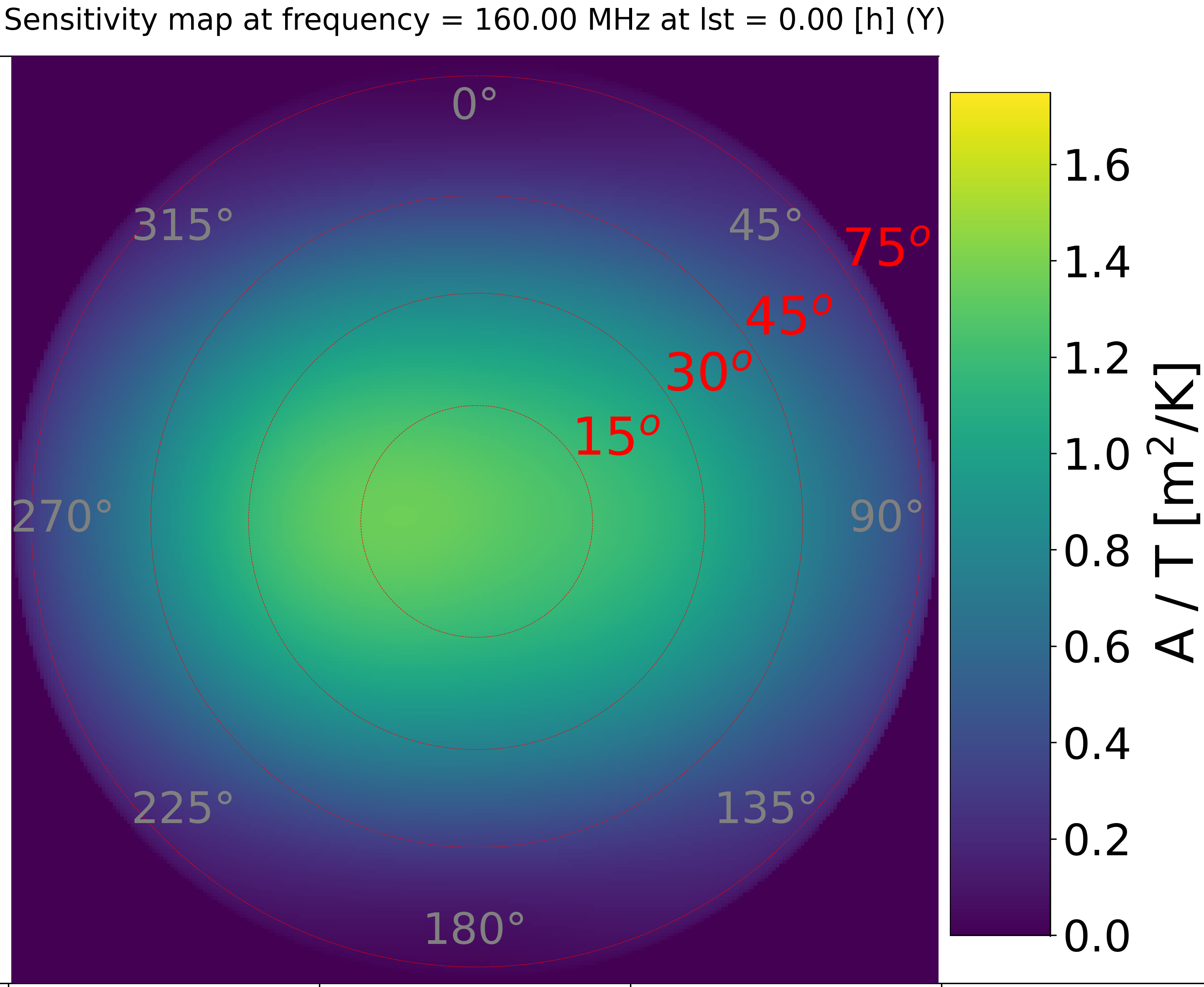} %
    \includegraphics[width=0.35\textwidth]{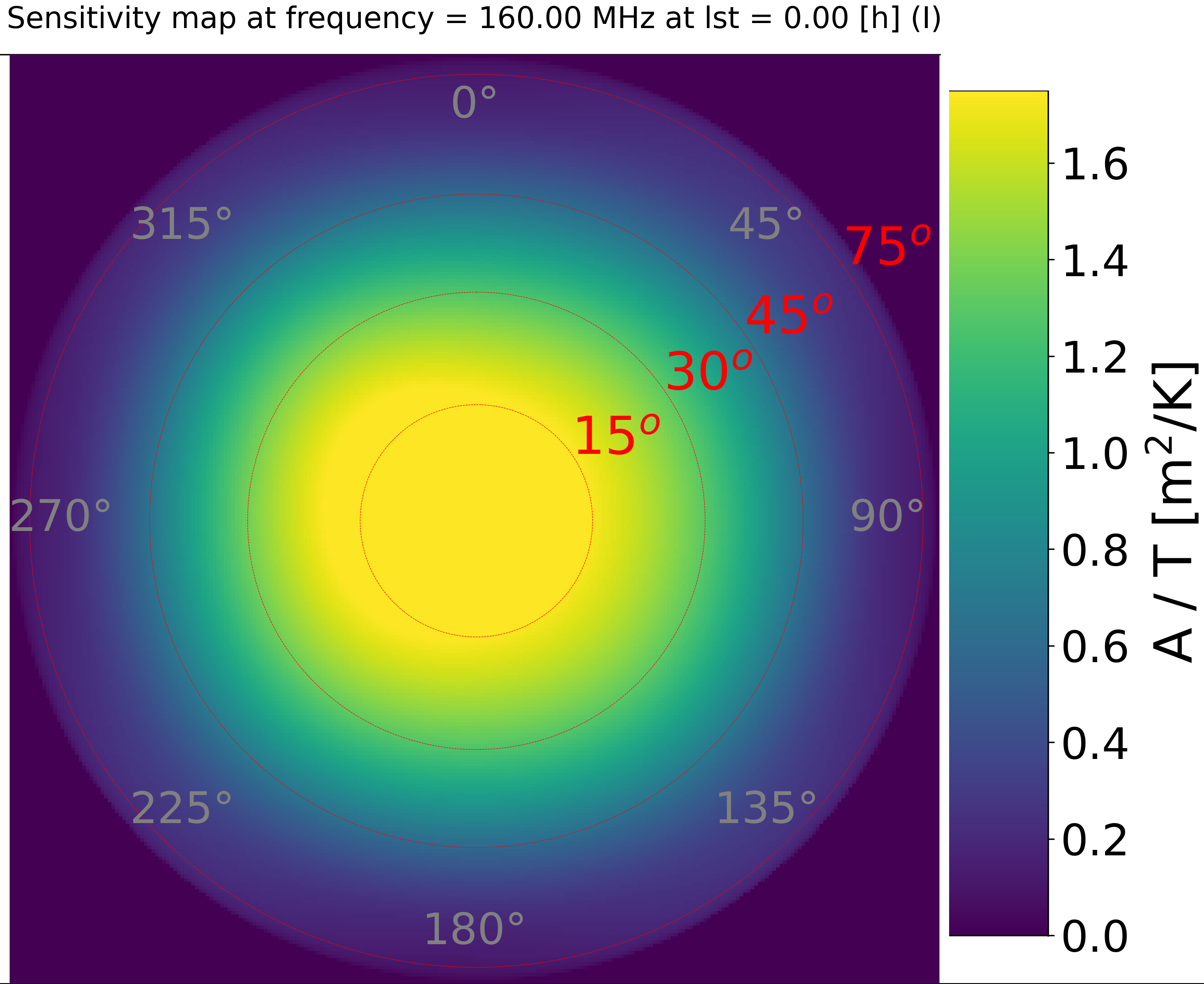} %
    \caption{The AAVS2 all-sky sensitivity map at frequency 160\,MHz and LST = 0\,h (``cold sky'') in X polarisation (left image), Y polarisation (centre image), and Stokes I polarisation (right image).}
    \label{fig_allsky_map_example_160mhz}
\end{figure*}

\begin{figure*}
   	\includegraphics[width=0.35\textwidth]{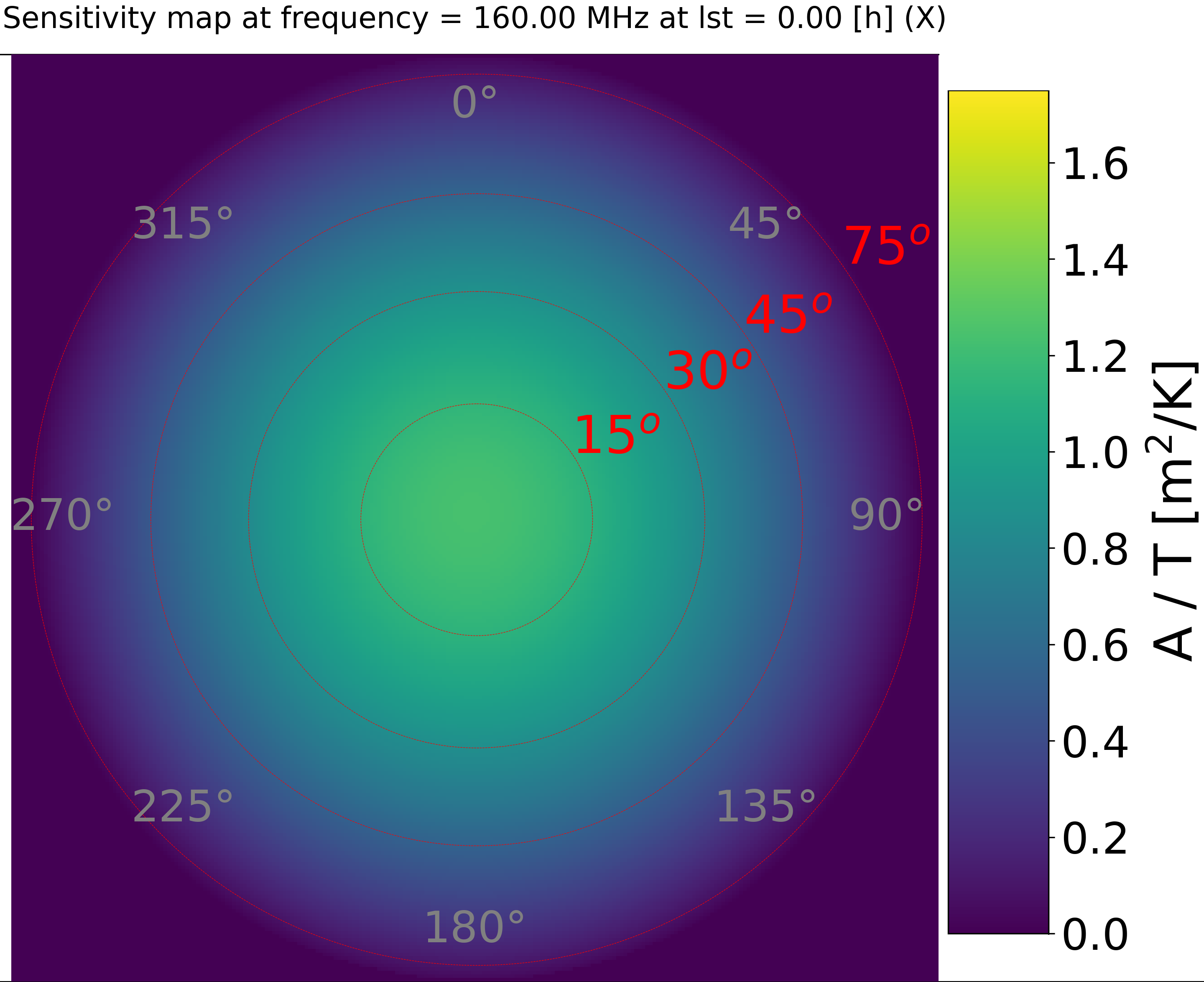} %
   	\includegraphics[width=0.35\textwidth]{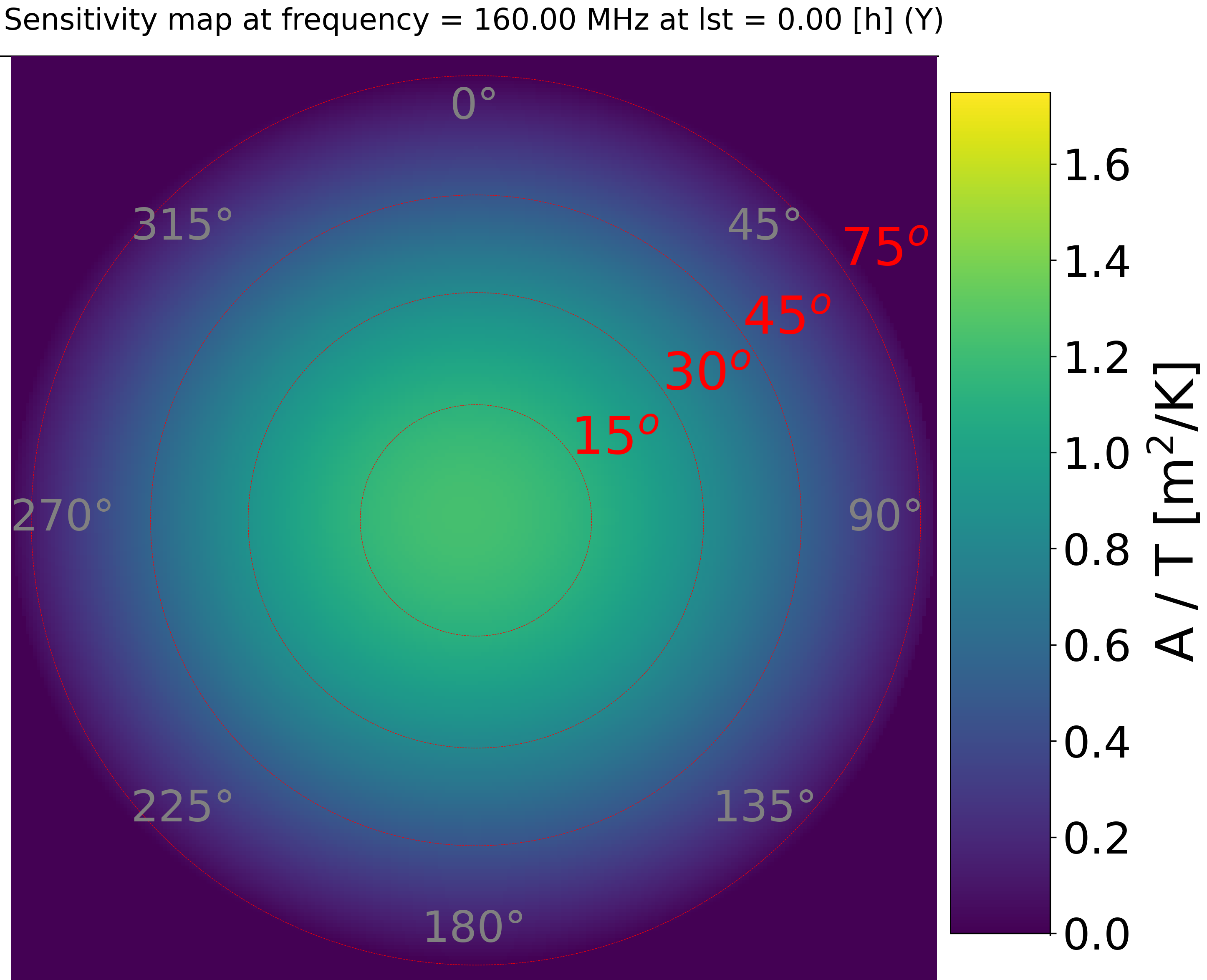} %
    \includegraphics[width=0.35\textwidth]{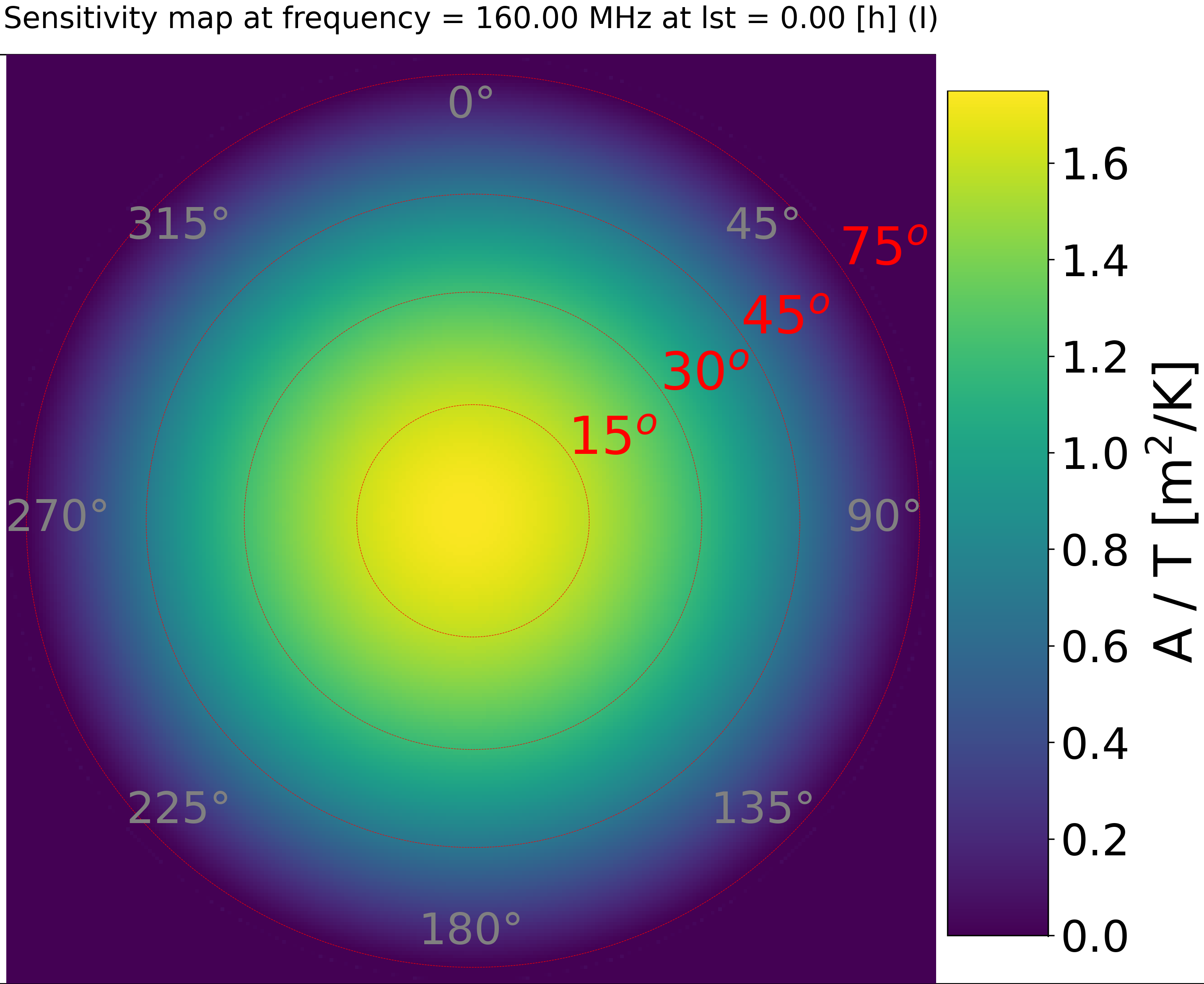} %
    \caption{The EDA2 all-sky sensitivity map at frequency 160\,MHz and LST = 0\,h (``cold sky'') in X polarisation (left image), Y polarisation (centre image), and Stokes I polarisation (right image).}
    \label{fig_allsky_map_example_160mhz_eda2}
\end{figure*}

\begin{figure*}
   	\includegraphics[width=0.35\textwidth]{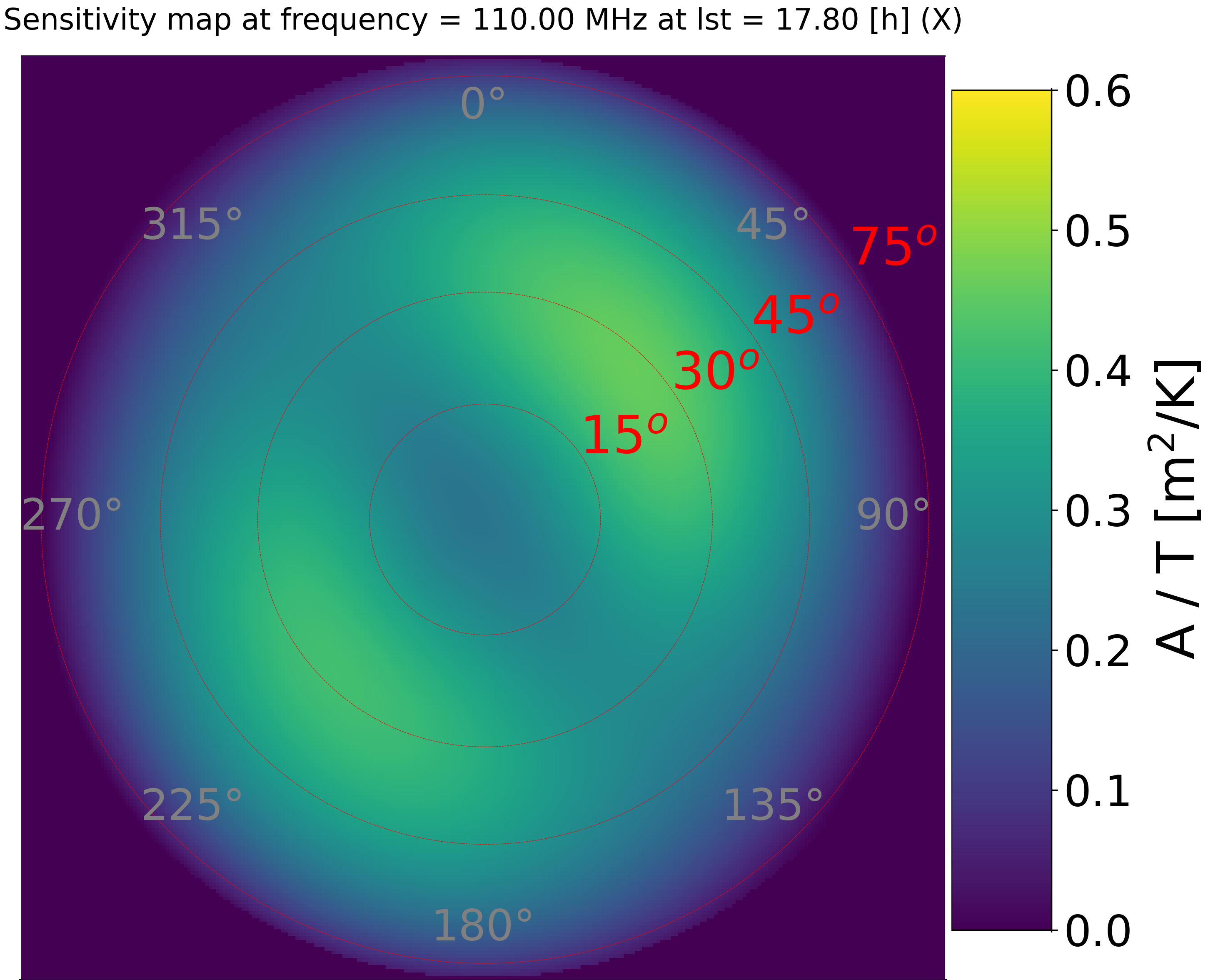} %
   	\includegraphics[width=0.35\textwidth]{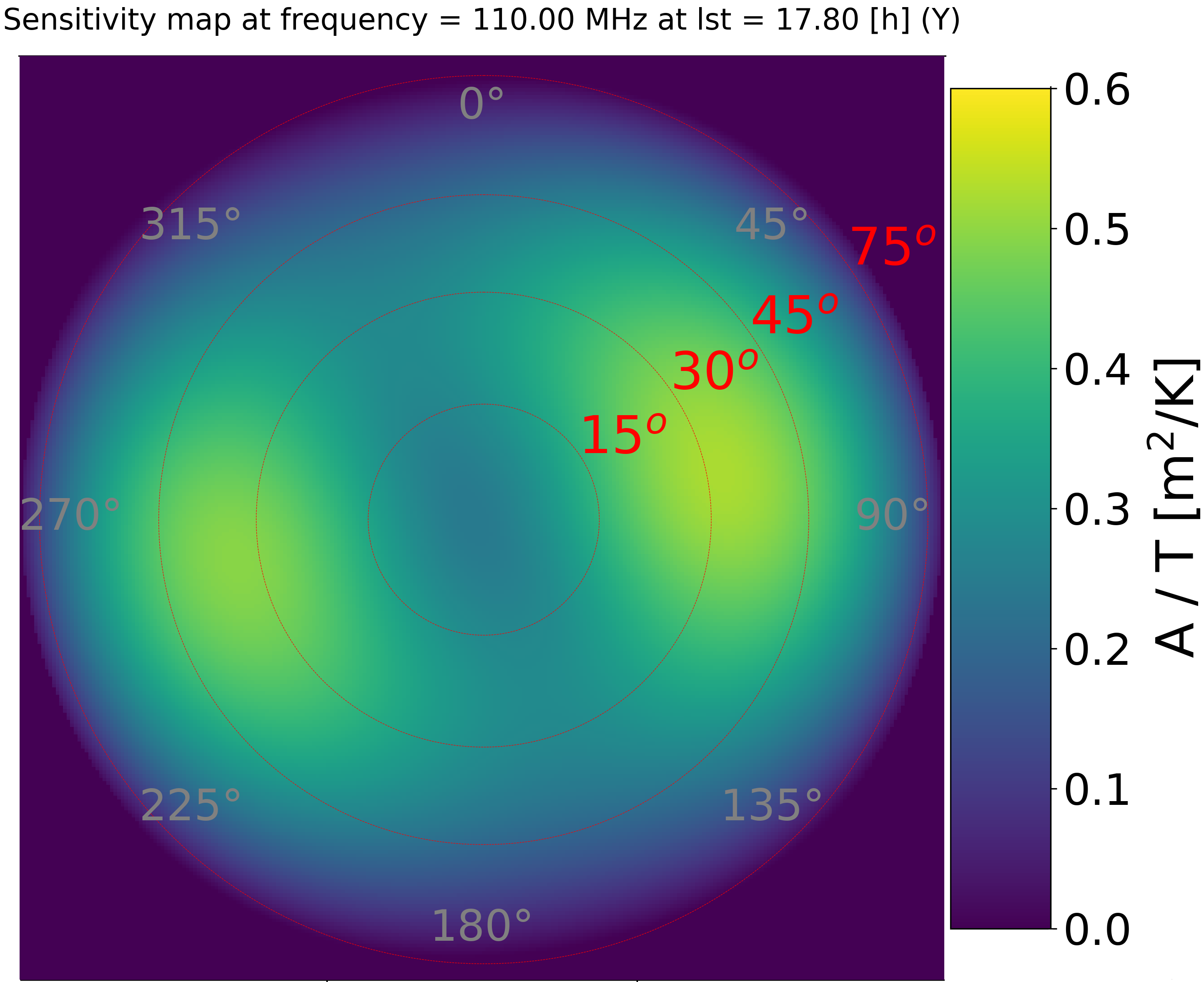} %
   	\includegraphics[width=0.35\textwidth]{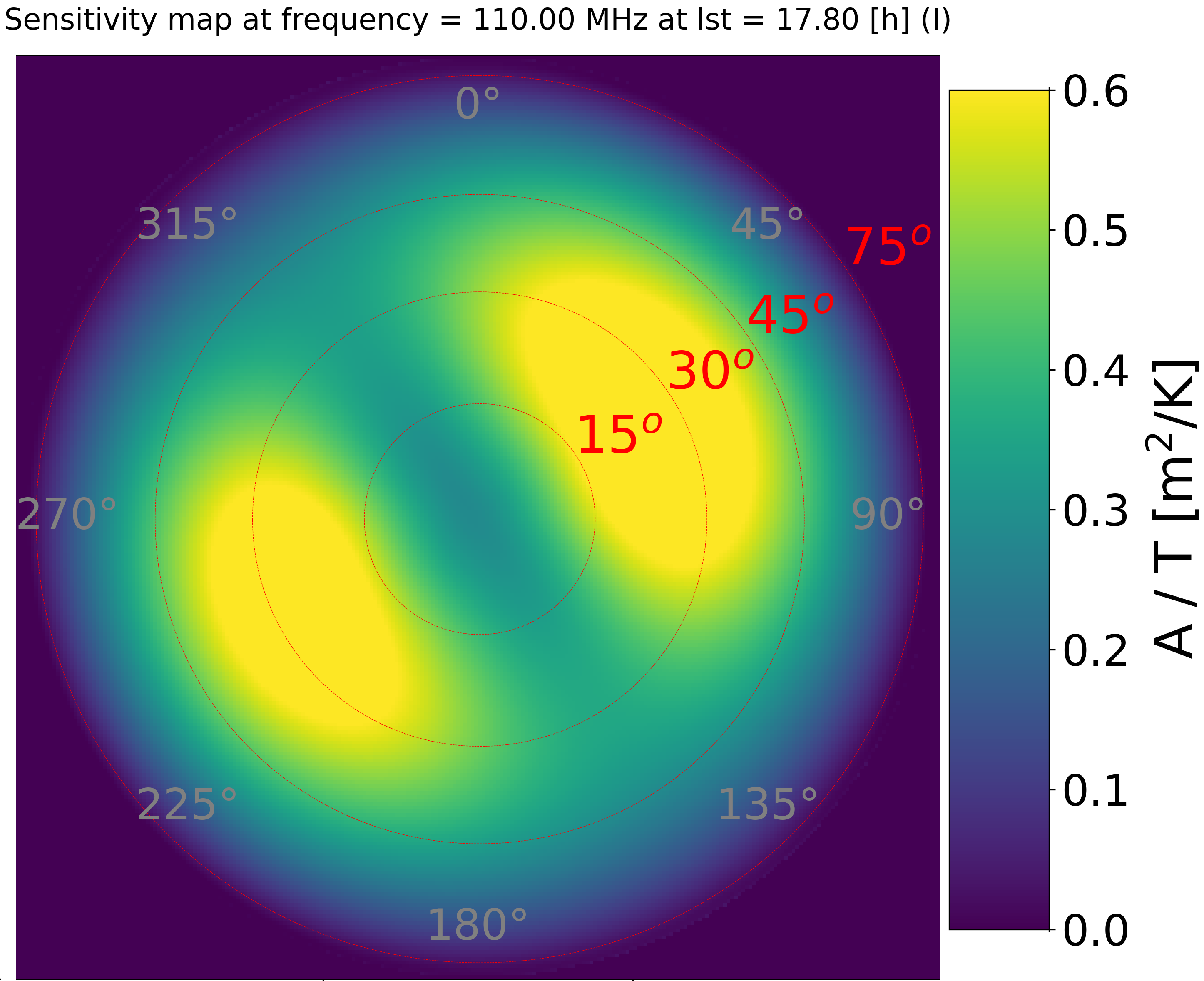} %
   	
   	\caption{\red{The AAVS2 all-sky sensitivity map at frequency 110\,MHz and LST = 17.8\,h (Galactic transit) in X polarisation (left image), Y polarisation (centre image), and Stokes I polarisation (right image). The clearly visible stripe of lower sensitivity is caused by high noise temperature at the Galactic Centre and Plane.}}
    \label{fig_allsky_map_example_110mhz_aavs2}
\end{figure*}

\begin{figure*}
   	\includegraphics[width=0.35\textwidth]{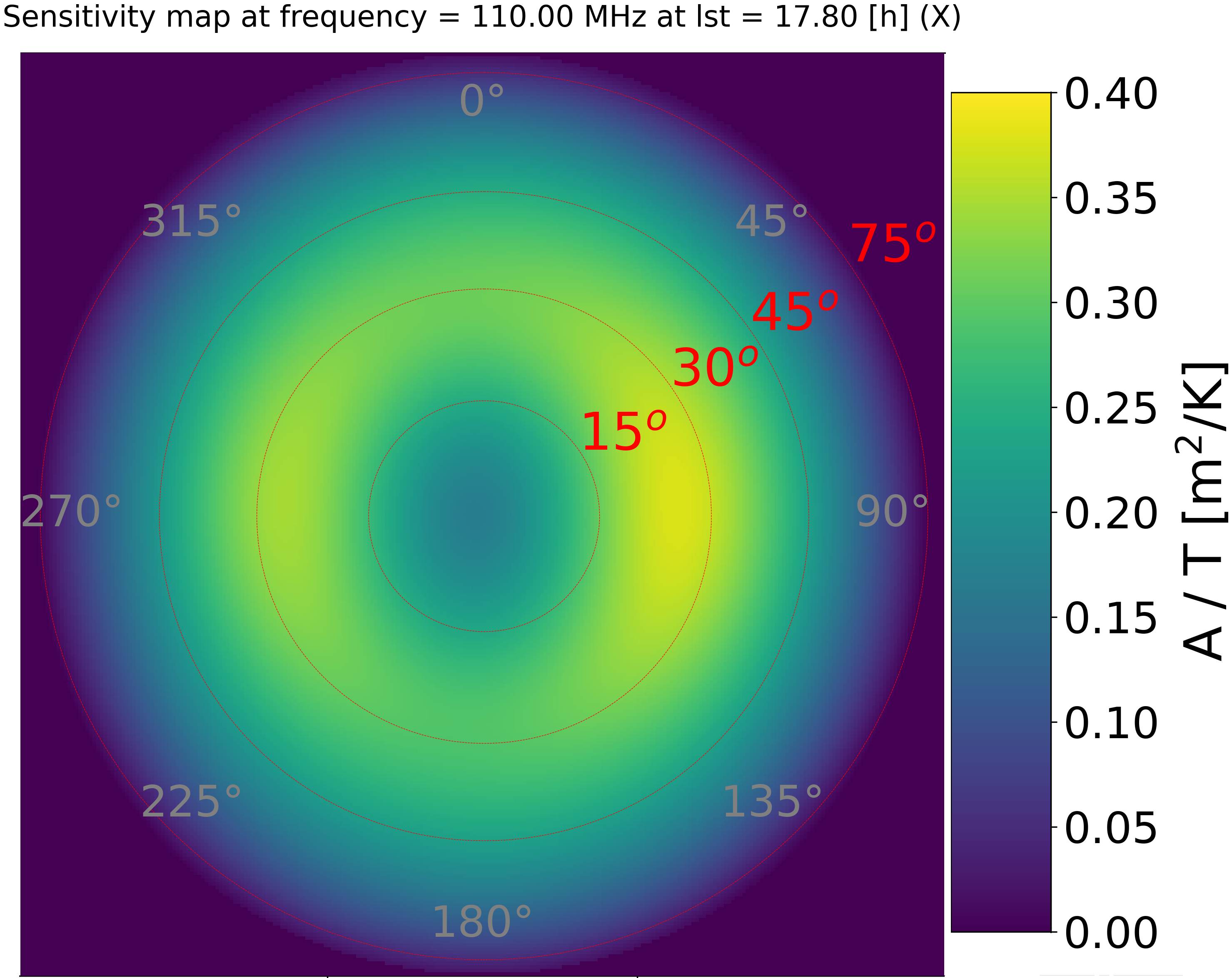} %
   	\includegraphics[width=0.35\textwidth]{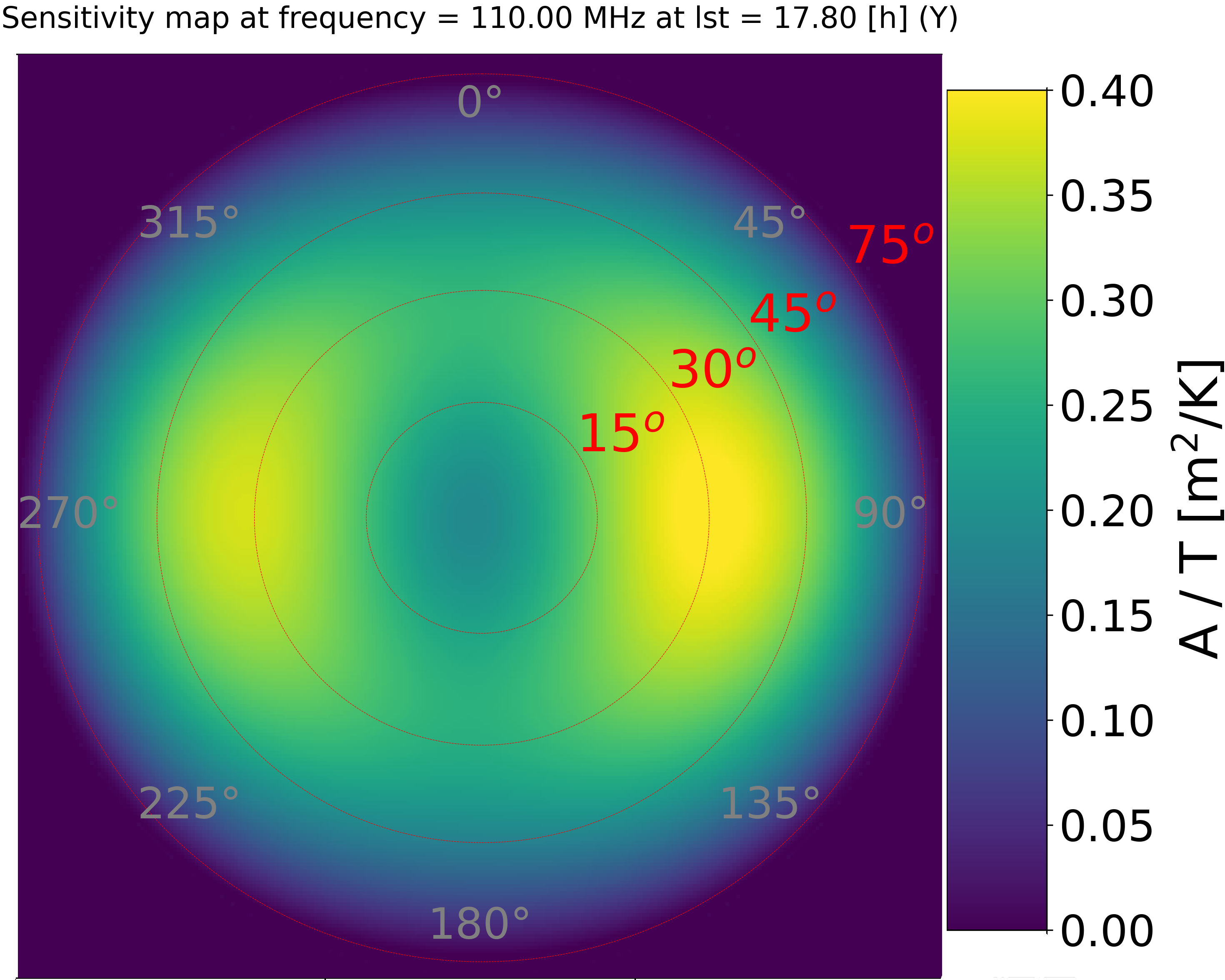} %
   	\includegraphics[width=0.35\textwidth]{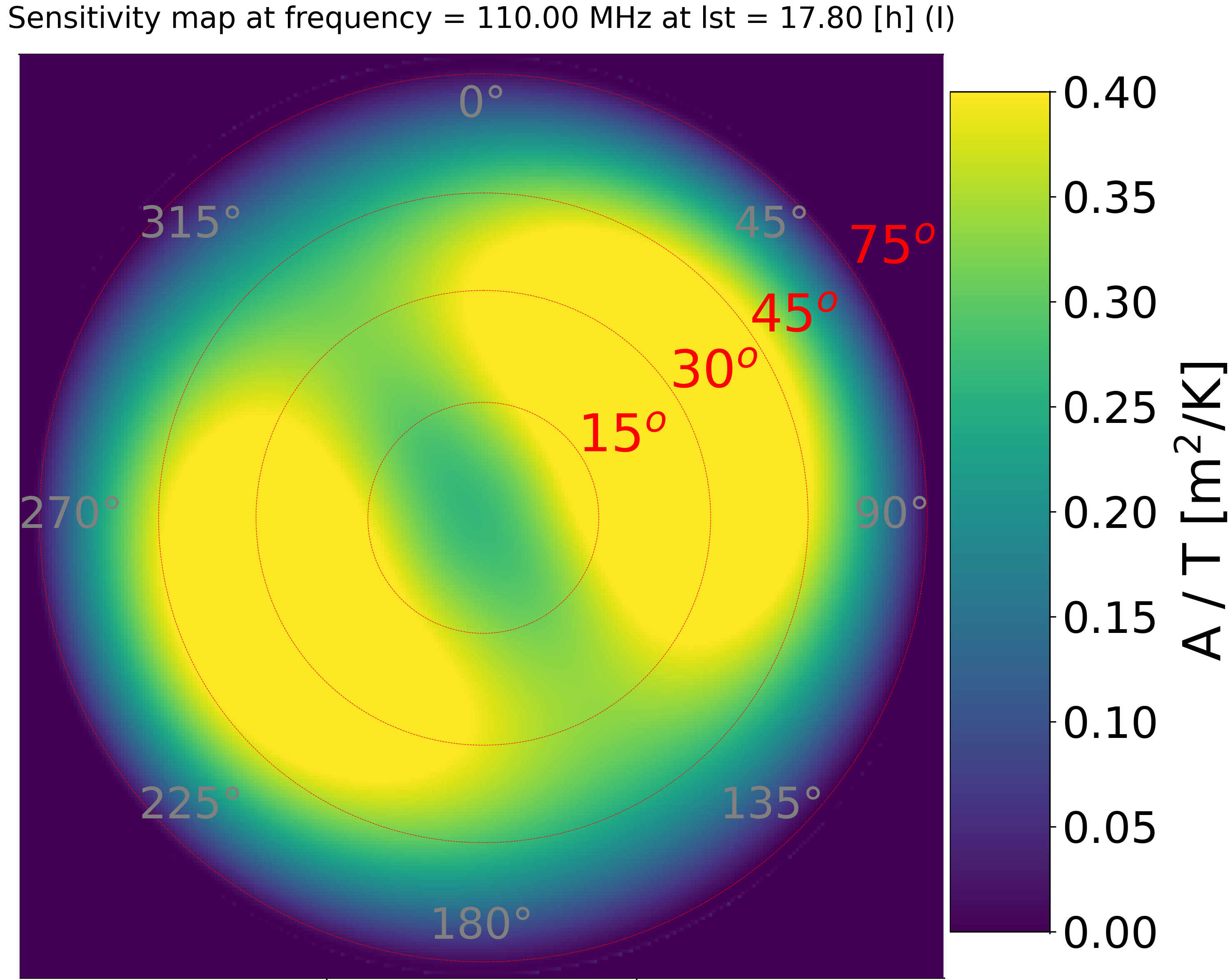} %
   	
   	\caption{\red{The EDA2 all-sky sensitivity map at frequency 110\,MHz and LST = 17.8\,h (Galactic transit) in X polarisation (left image), Y polarisation (centre image), and Stokes I polarisation (right image). The clearly visible stripe of lower sensitivity is caused by high noise temperature at the Galactic Centre and Plane.}}
    \label{fig_allsky_map_example_110mhz_eda2}
\end{figure*}

\begin{figure*}
   	\includegraphics[width=0.35\textwidth]{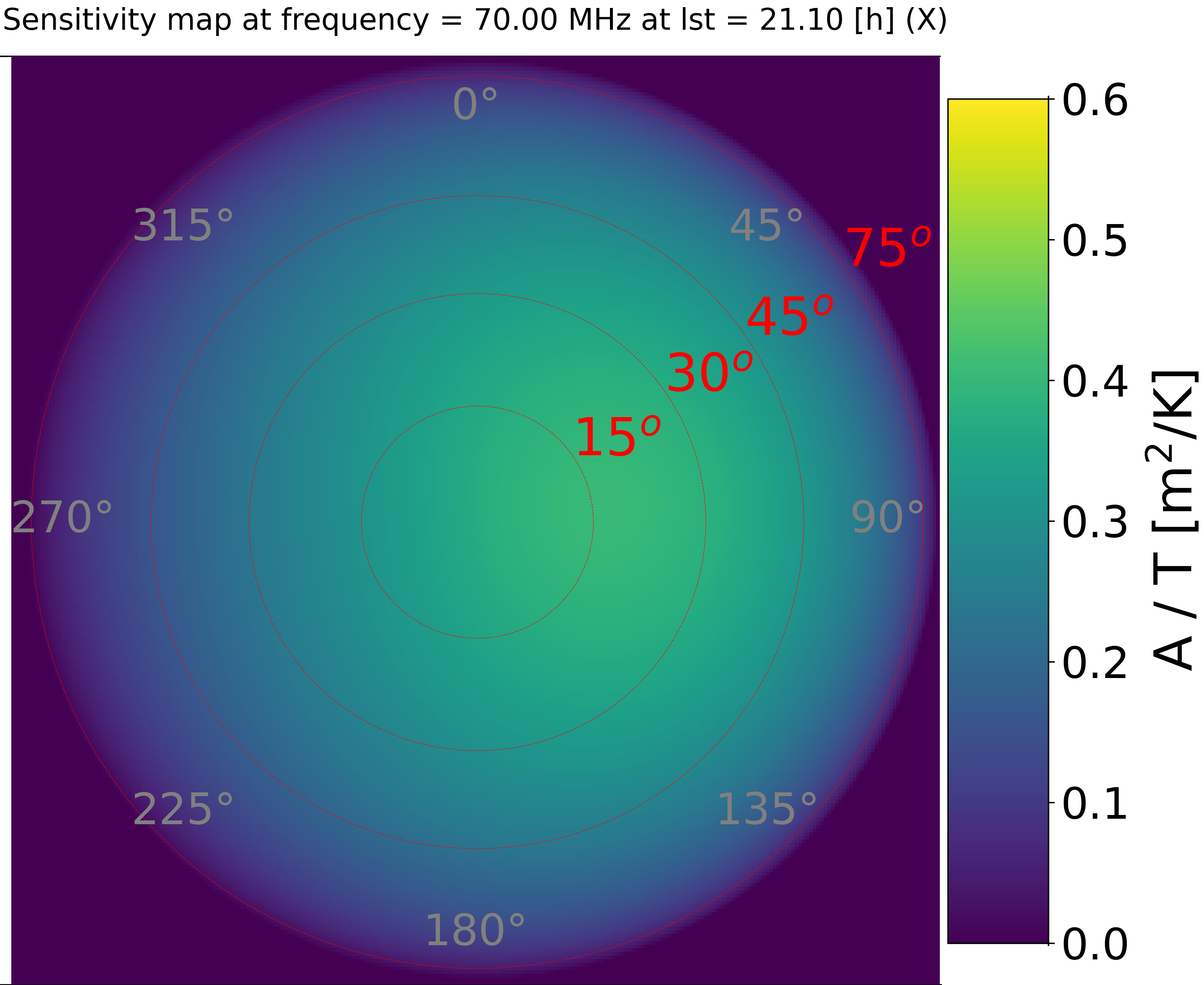} %
   	\includegraphics[width=0.35\textwidth]{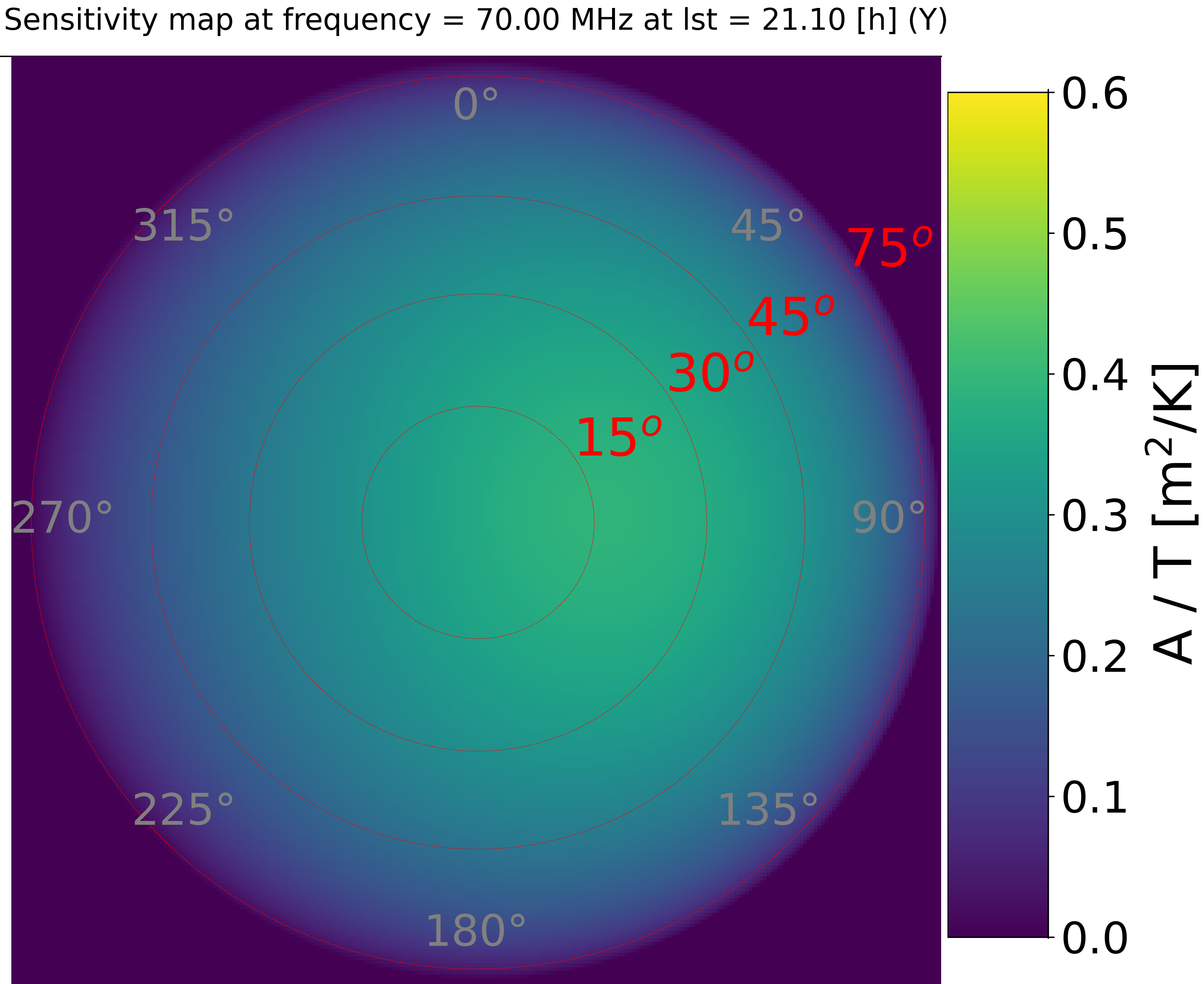} %
   	\includegraphics[width=0.35\textwidth]{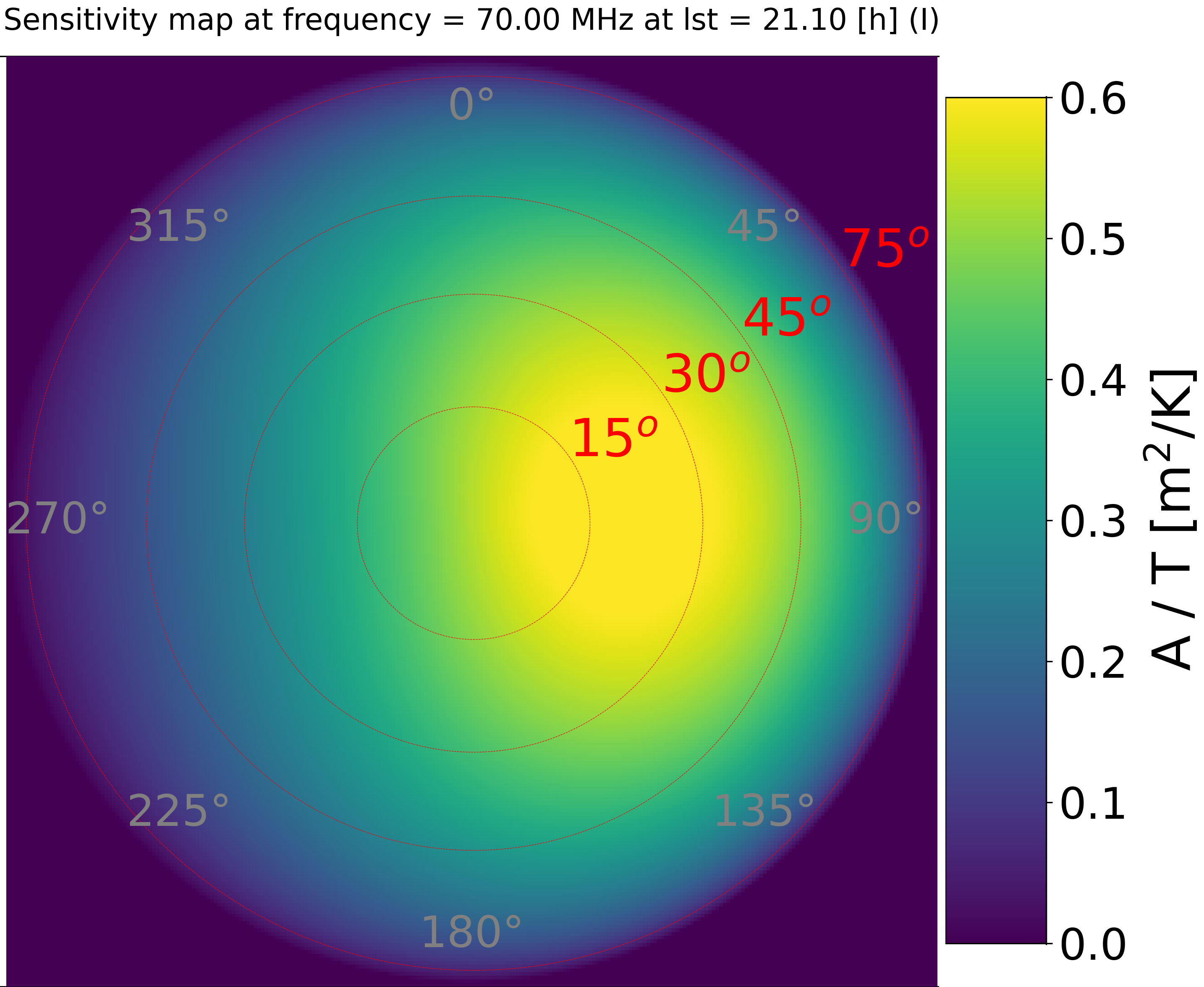} %
   	
   	\caption{The AAVS2 all-sky sensitivity map at frequency 70\,MHz and LST = 21.1\,h (Galactic center at the elevation $\approx$45\degree\, in the West at azimuth 270\degree) in X polarisation (left image), Y polarisation (centre image), and Stokes I polarisation (right image).}
    \label{fig_allsky_map_example_70mhz}
\end{figure*}

\begin{figure*}
   	\includegraphics[width=0.35\textwidth]{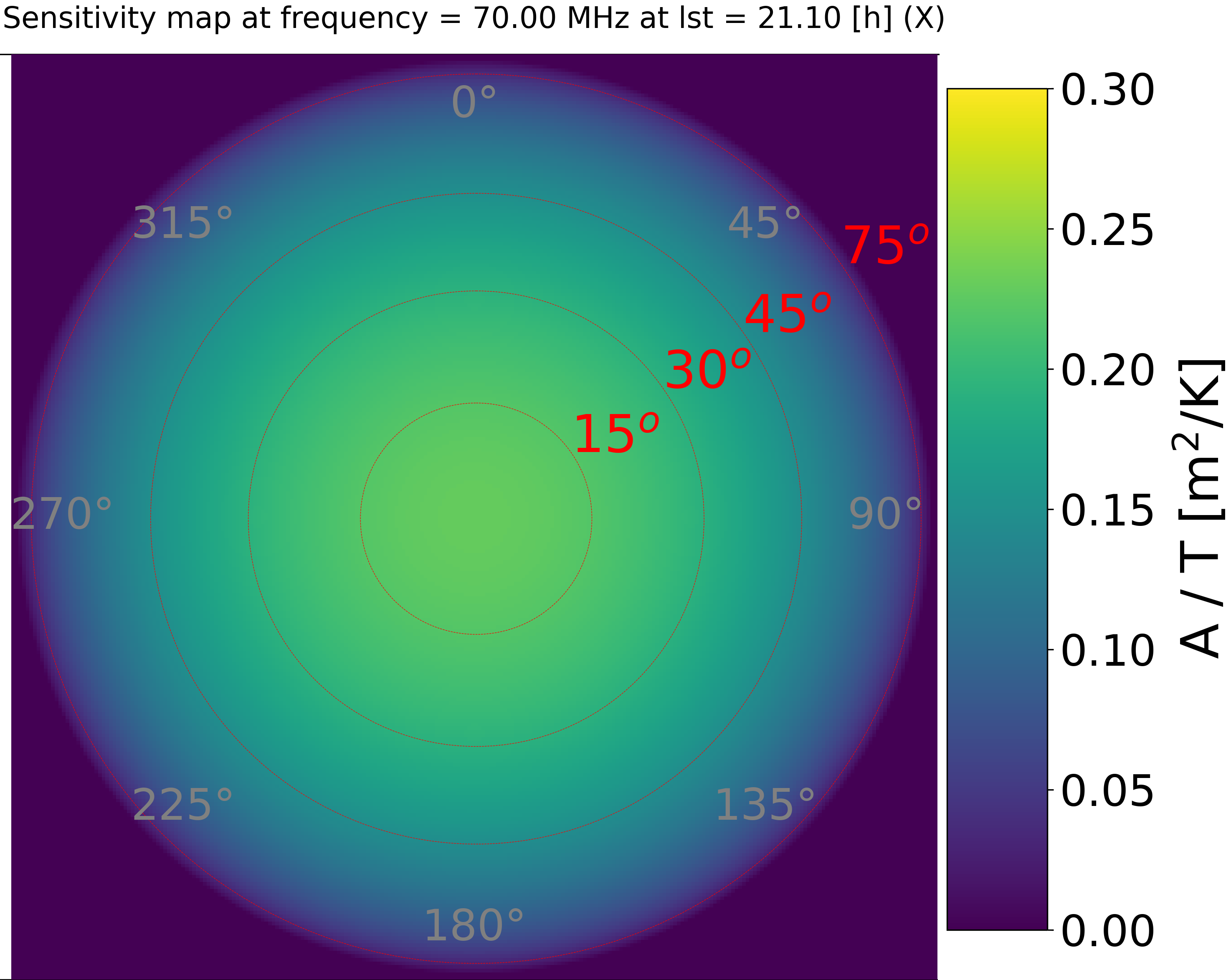} %
   	\includegraphics[width=0.35\textwidth]{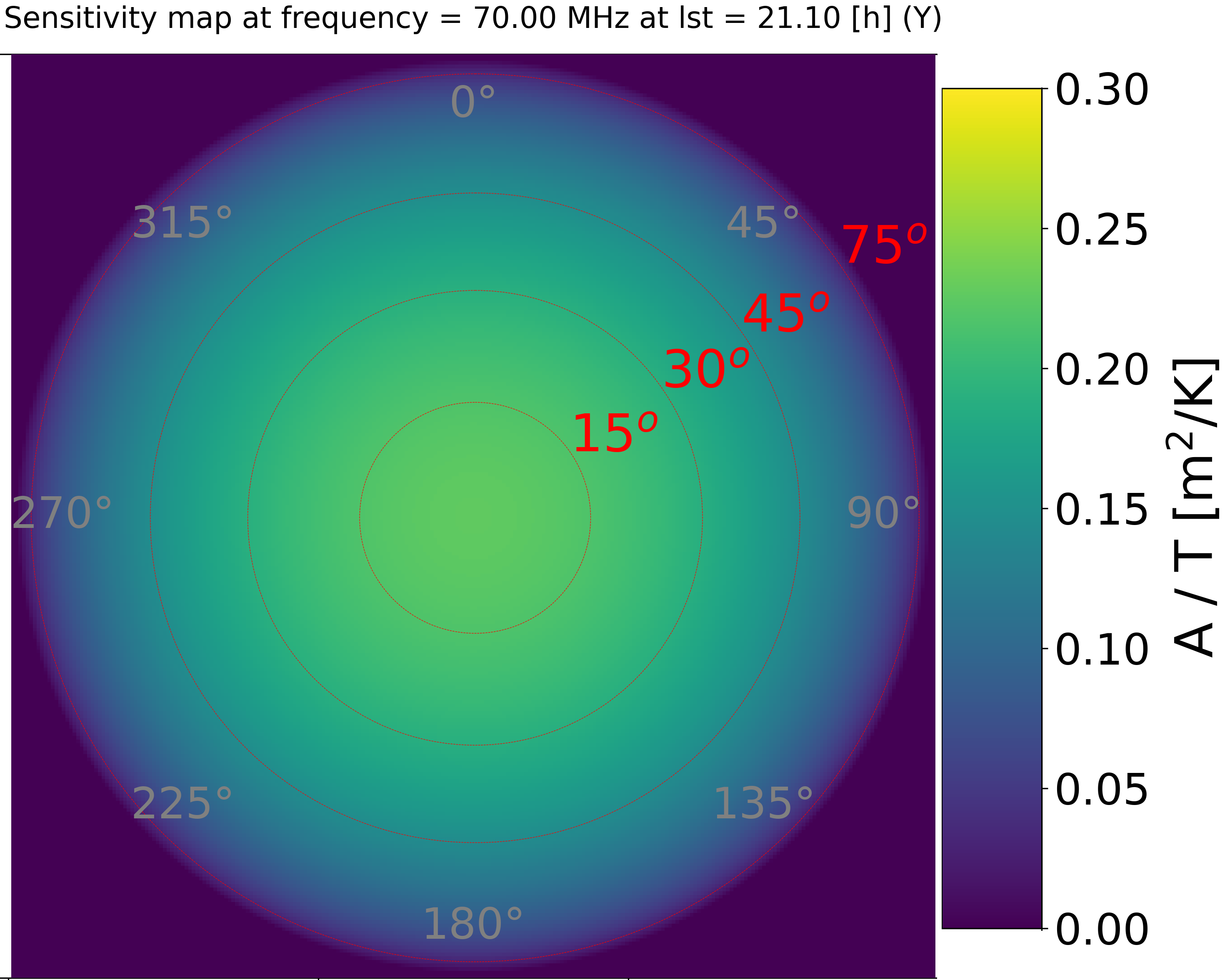} %
   	\includegraphics[width=0.35\textwidth]{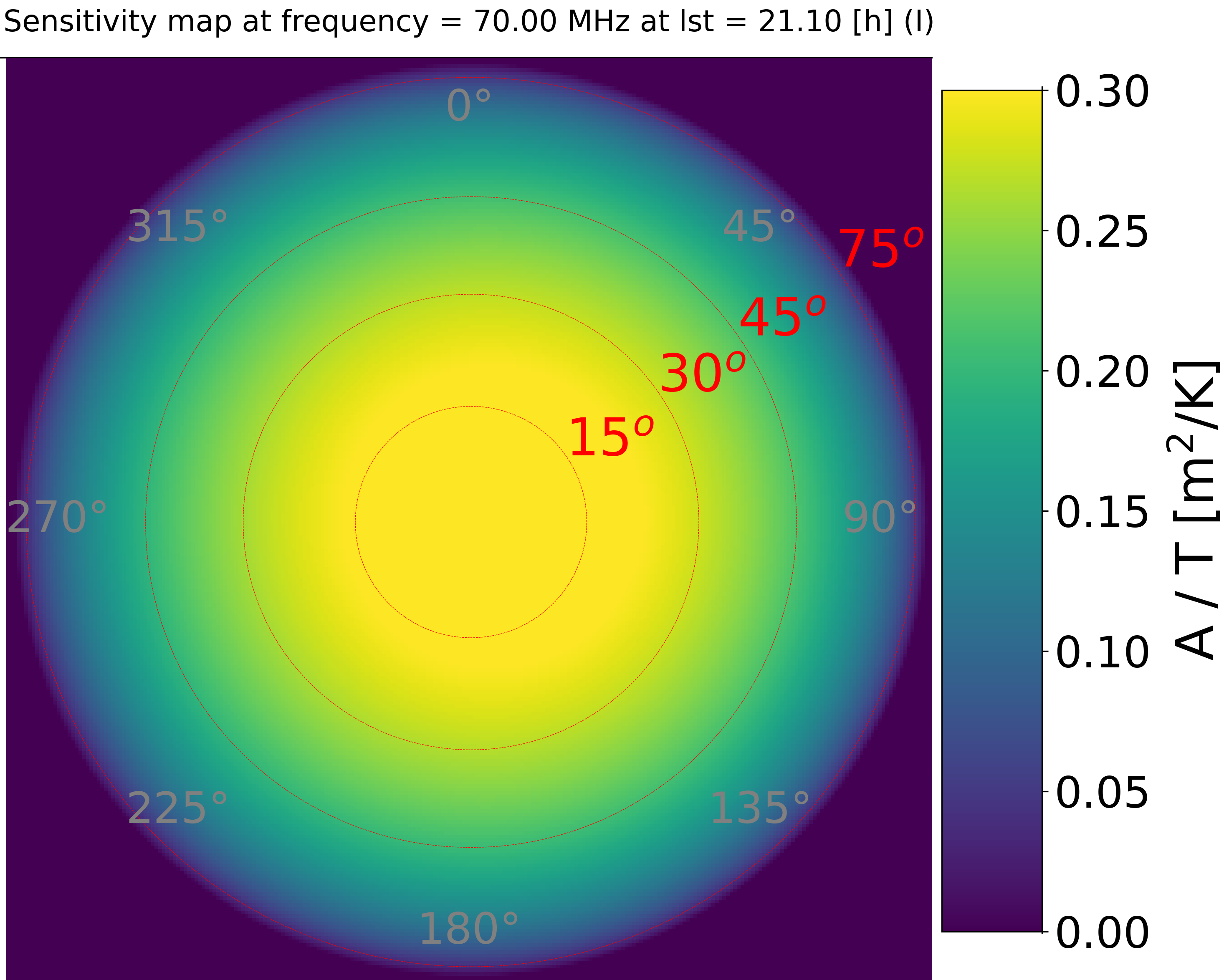} %
   	
   	\caption{The EDA2 all-sky sensitivity map at frequency 70\,MHz and LST = 21.1\,h (Galactic center at the elevation $\approx$45\degree\, in the West at azimuth 270\degree) in X polarisation (left image), Y polarisation (centre image), and Stokes I polarisation (right image).}
    \label{fig_allsky_map_example_70mhz_eda2}
\end{figure*}

\begin{figure*}
   	\includegraphics[width=\columnwidth]{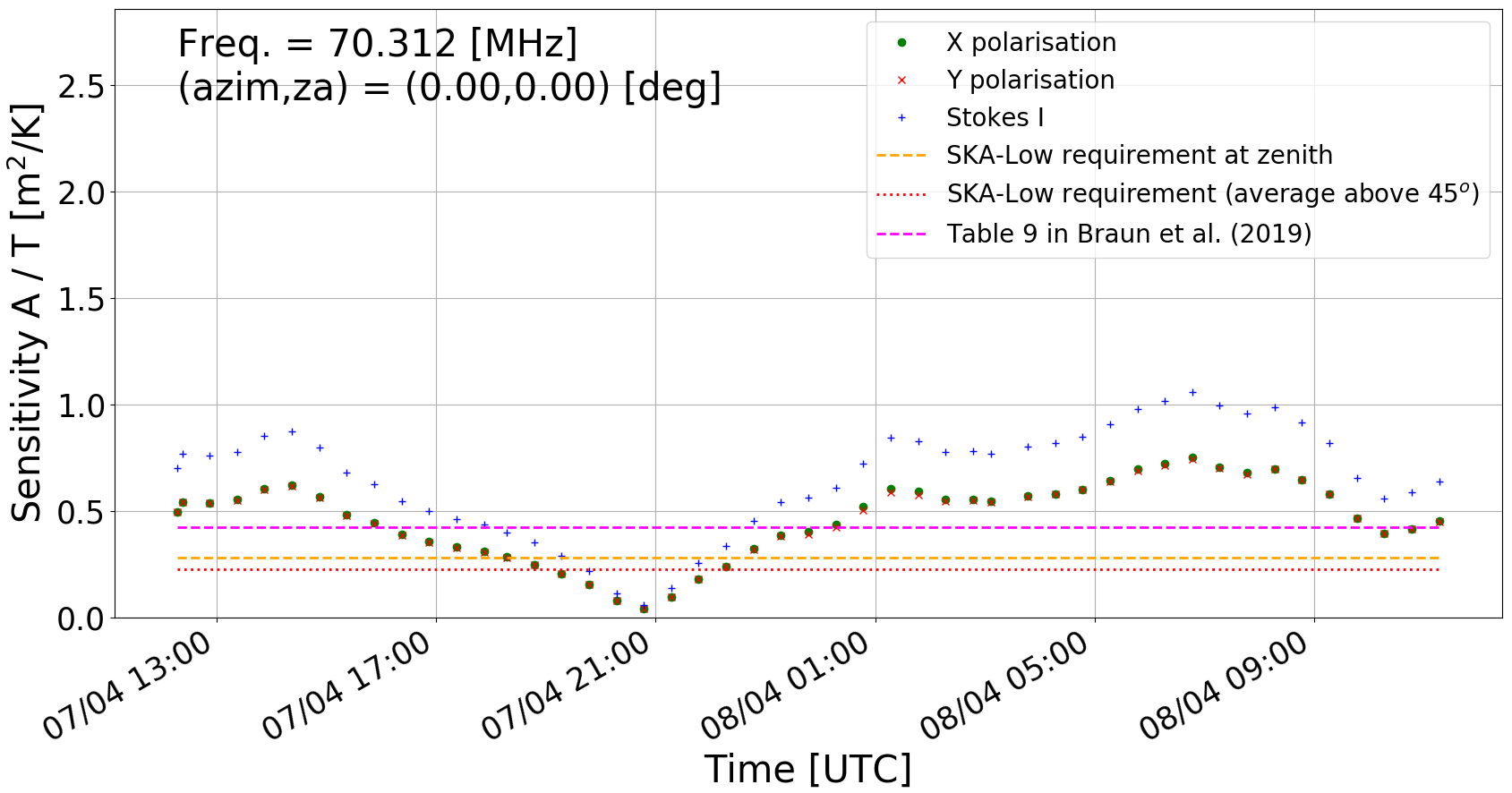} %
   	\includegraphics[width=\columnwidth]{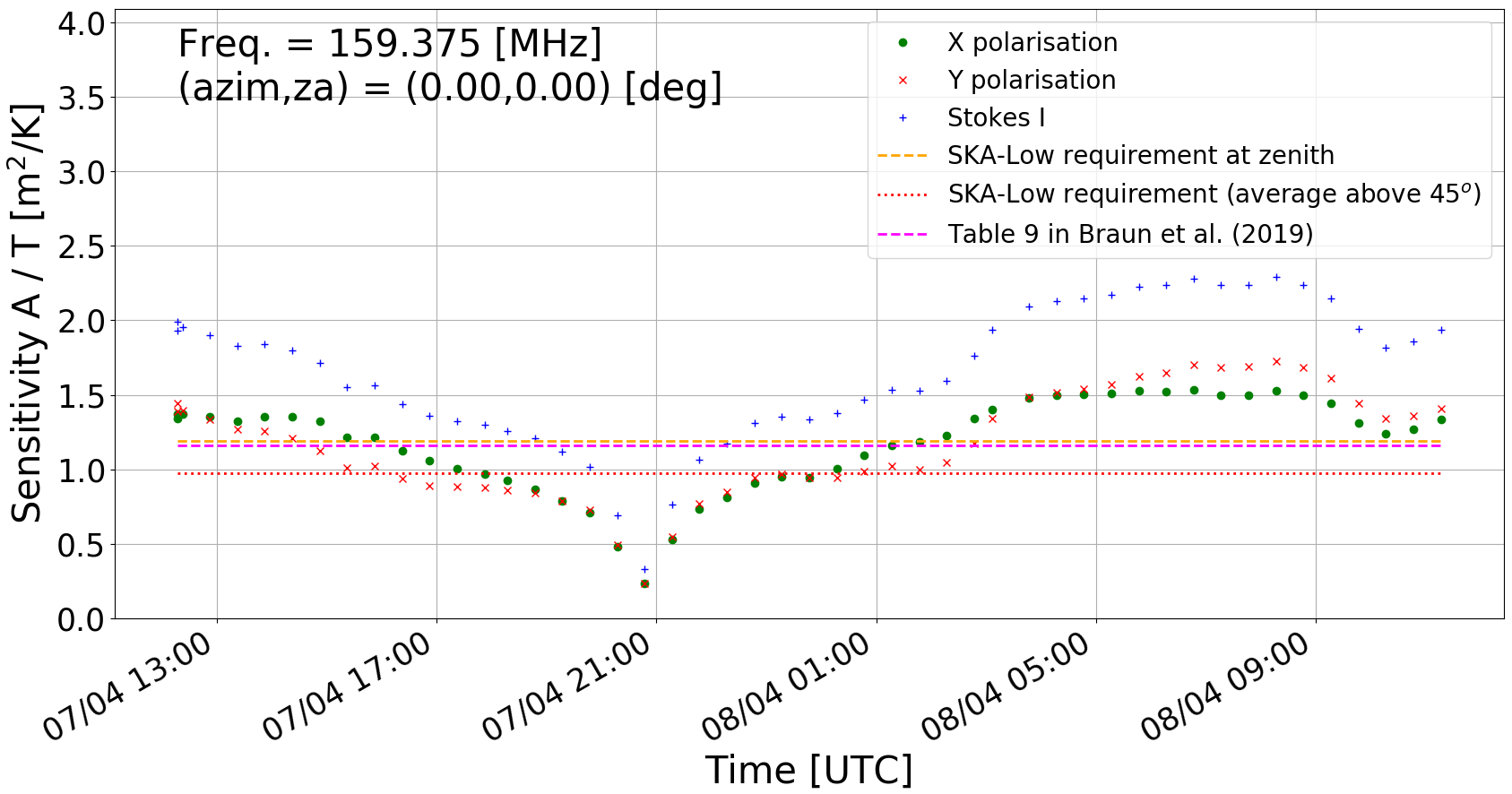} %
   	
    \caption{The AAVS2 sensitivity as a function of time at the zenith pointing. Left image : frequency 70.3125\,MHz. Right image : frequency 159.375\,MHz. These data were generated with a station beam model ($B_{st}(\nu,\theta,\phi)$) in equation~\ref{eq:sky_integration}. Noticeable is the sharp drop in sensitivity at time of the Galactic transit, which is caused by a very sharp peak in antenna temperature (as in Figure~\ref{fig_aavs2_drift_scan_ch204}) causing sharp and significant reduction in sensitivity. \red{The data points are results of this work, dashed orange and dotted red curves are SKA-Low specifications at zenith and averaged over elevations $\ge$45\degree \,, respectively. The magenta dotted curve is the sensitivity from Table 9 in \citet{2019arXiv191212699B} (also averaged over elevations $\ge$45\degree).}}
    \label{fig_aavs2_sens_vs_time_azim0_za0}
\end{figure*}

\begin{figure*}
   	\includegraphics[width=\columnwidth]{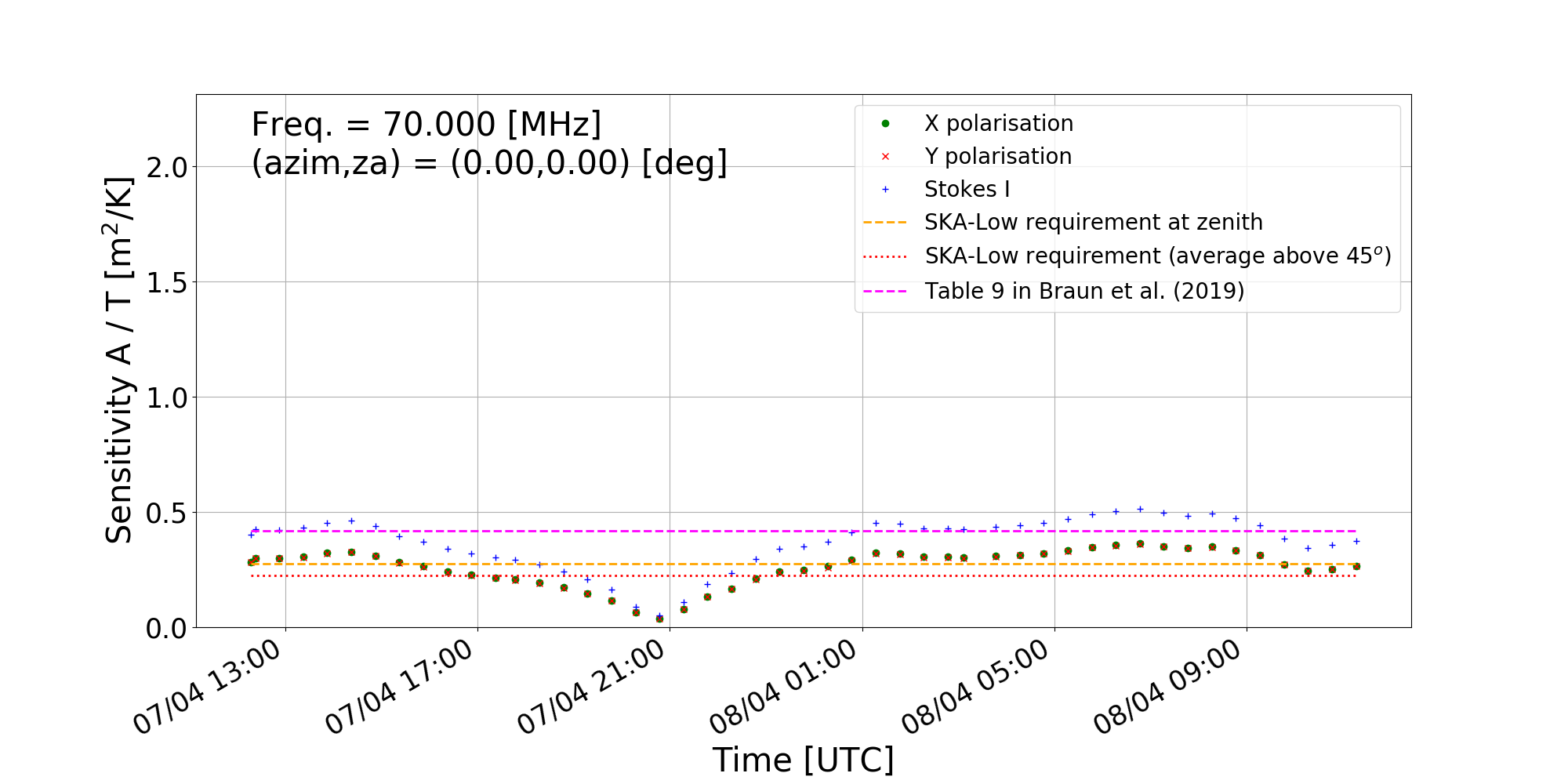} %
   	\includegraphics[width=\columnwidth]{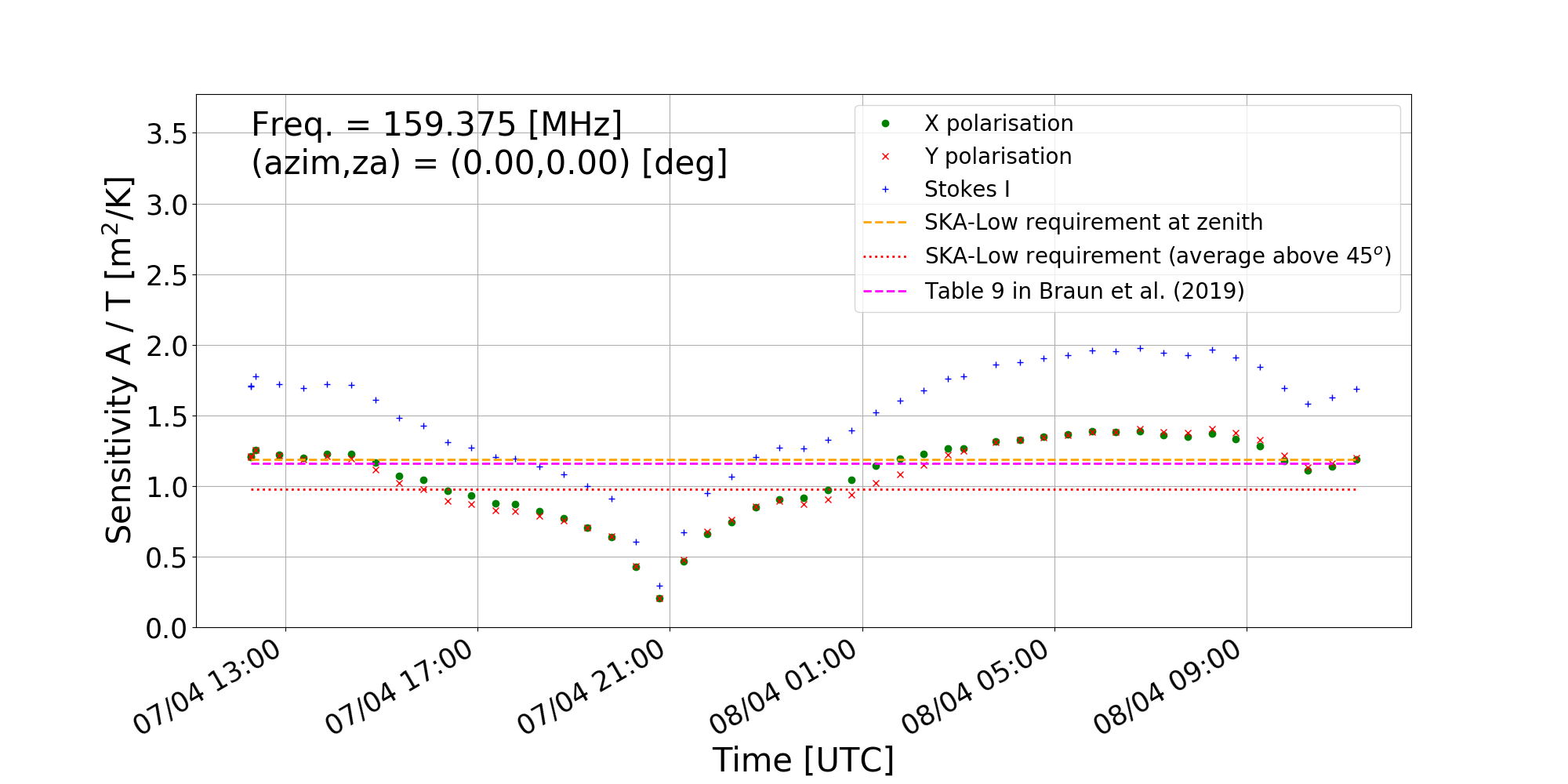} %
   	
    \caption{The EDA2 sensitivity as a function of time at the zenith pointing. Left image : frequency 70.3125\,MHz. Right image : frequency 159.375\,MHz. These data were generated with a station beam model ($B_{st}(\nu,\theta,\phi)$) in equation~\ref{eq:sky_integration}. Noticeable is the sharp trough in sensitivity at time of the Galactic transit, which is caused by a very sharp peak in antenna temperature (as in Figure~\ref{fig_aavs2_drift_scan_ch204} for AAVS2) causing sharp and significant reduction in sensitivity. \red{The data points are results of this work, dashed orange and dotted red curves are SKA-Low specifications at zenith and averaged over elevations $\ge$45\degree \,, respectively. The magenta dotted curve is the sensitivity from Table 9 in \citet{2019arXiv191212699B} (also averaged over elevations $\ge$45\degree).}}
    \label{fig_eda2_sens_vs_time_azim0_za0}
\end{figure*}

The \textsc{python} script \texttt{sensitivity\_db.py} enables access to the sensitivity database. This script requires \textsc{SQLite} database files to exist locally. The full deployment procedure is described in the \textsc{github} repository. On execution, the script generates output \textsc{text} files and \textsc{png} images according to the request specified by the command line options. Typical command line parameters are described in the user manual in the \textsc{github} repository and Appendix~\ref{appendix_cmdline_options}; they implement the following functionality:

\begin{itemize}
   \item \textbf{Sensitivity as a function of frequency} at a specified pointing direction and LST or UTC time. Example images obtained with this option at the LST$=0$\,h (``cold sky'' transiting overhead) and at LST$=17.8$\,h (Galactic Centre transit) are shown respectively in the left and right panels of Figures~\ref{fig_aavs2_sens_vs_freq_azim0_za0} and \ref{fig_eda2_sens_vs_freq_azim0_za0} for AAVS2 and EDA2 respectively. \red{For comparison, SKA-Low specifications and sensitivity from Table 9 in \citet{2019arXiv191212699B} were included in these plots showing that at some LSTs they can be significantly different (by even a factor of a few) from the results of our simulations.}\\
   
   \item \textbf{All-sky sensitivity map} at a specified frequency and LST or UTC time. Example images obtained with this option at LST$=0$\,h (``cold sky'') and frequency 159.375\,MHz are shown in Figures~\ref{fig_allsky_map_example_160mhz} and ~\ref{fig_allsky_map_example_160mhz_eda2} for AAVS2 and EDA2 respectively. \red{Example images obtained with this option at the time of Galactic transit LST$=17.8$\,h (``hot sky'') and frequency 110\,MHz are shown in Figures~\ref{fig_allsky_map_example_110mhz_aavs2} and~\ref{fig_allsky_map_example_110mhz_eda2} for AAVS2 and EDA2 respectively. These images clearly demonstrate significant non-uniformity of the sensitivity across the sky reaching even a factor of 3 between zenith (high sky noise temperature at the Galactic Centre and Plane) and some 30\degree \,away from zenith in the North-East direction. } Corresponding images at LST$=21$\,h (Galactic Centre in the West at elevation $\approx$45\degree) and frequency 70.3125\,MHz are presented in Figures~\ref{fig_allsky_map_example_70mhz} and ~\ref{fig_allsky_map_example_70mhz_eda2} for AAVS2 and EDA2 respectively.\\ 
   
   \item \textbf{Sensitivity as a function of time} at a specified pointing direction and frequency. Example images obtained with this option at the zenith pointing and frequencies 70.3125 and 159.375\,MHz are shown in the left and right panels of Figures~\ref{fig_aavs2_sens_vs_time_azim0_za0} and \ref{fig_eda2_sens_vs_time_azim0_za0} for AAVS2 and EDA2 respectively. \red{For comparison, SKA-Low specifications and sensitivity from Table 9 in \citet{2019arXiv191212699B} were also included. Both are time independent, while the values resulting from our work vary significantly with LST (nearly by an order of magnitude) resulting in even factor $\sim$2 differences with the work by \citet{2019arXiv191212699B} and SKA-Low specifications.}\\
\end{itemize}

In addition to generating images, the above options can also save the results to \textsc{text} files, which can be used in later analysis.

\subsection{Web interface}
\label{subsec:web_interface}

The same functionality as provided by the command line script described in Section~\ref{subsec:command_line} is also provided by the web interface. This interface has been implemented as a \textsc{Django} web service deployed on the Amazon web-service available at \skacalcwebpage. The user can specify a station name (presently EDA2 or AAVS2) and request sensitivity data in one of the formats described in Section~\ref{subsec:command_line}, and the system will generate and display plots of sensitivity as a function of requested parameters (these images can be saved as \textsc{png} files). Optionally, the user can also request \textsc{text} files with the sensitivity data and in such a case a \textsc{zip} archive file containing the requested data in both \textsc{text} and \textsc{png} formats will be returned and can be saved on the user's local hard-drive for later analysis.  

\section{Verification of the sensitivity simulations using astronomical observations}
\label{sec:eda2_aavs2_comparison}

Predictions of the sensitivity simulations were verified using astronomical data. The sensitivity at zenith was measured from difference images of the entire hemisphere and compared against sensitivity simulations generated at the corresponding frequency, time range, and zenith pointing. These comparisons were performed and published earlier as a part of verification of stations performance and sensitivity. The details of the observations, data processing, and comparisons between the data and simulations for the AAVS2 station can be found in \citet{9410962}, while the comparison analysis for EDA2 has been published in \citep{eda2_paper}. Both these analyses used our station simulation package to calculate the sensitivity of both stations, which simultaneously observed the same frequency channel (to enable direct comparisons). We note that further analysis of the AAVS2 sensitivity is also described by \citet{giulia_et_al} but a different simulation package was used in their analysis.

The astronomical data used in the verification were collected in April, May, and September 2020 at a single frequency channel of approximately 0.94\,MHz width. Each dataset spans a time interval of at least 24\,h and the data were phase and flux density calibrated using transiting Sun observations and a quiet Sun model \citep{sun_model} as a calibrator. The calibration of each dataset was performed at a solar transit and the resulting calibration solutions were applied across the entire dataset. Therefore, the agreement between the data and the simulations is always the best (within 10\%) at the calibration time (near local midday). The discrepancy between the data and simulations increases slightly (up to 20-30\%) away from the calibration time and this was found to be caused by the gain variations resulting from diurnal changes in ambient temperature. Nevertheless, \citet{eda2_paper} used equation~\ref{eq:sky_integration} (this paper) with the model of the station beam ($B_{\text{st}}$) replaced by the beam model of a single dipole ($B_{\text{dip}}$) to calculate expected total power from a single antenna as a function of time ($\text{S}_\text{dip}(\text{LST})$). Then the measured power from each individual dipole ($\text{P}_\text{dip}(\text{LST})$)  was divided by the simulated power ($\text{S}_\text{dip}(\text{LST})$) to calculate the gain of each individual antenna as a function of time, $\text{G}_\text{dip}(\text{LST}) = \text{P}_\text{dip}(\text{LST}) / \text{S}_\text{dip}(\text{LST})$, which resulted in 256 gain curves for each of the polarisations (X and Y). Finally, the median gain curve was calculated as a median (at each timestep) of all 256 curves and normalised by the gain value at the calibration time. After multiplying the sensitivity data (SEFD) by this normalised gain, the agreement between the sensitivity data and the simulations is excellent (e.g. Figures 10 and 11 in \citet{eda2_paper}). 

These verifications give us confidence that the array factor method is accurate to within a few percent at zenith, which was further confirmed in simulations by \citet{davids_paper} showing that the inaccuracies of the simulations increase only at elevations below 50\degree. The inaccuracy of calculating the SEFD of the Stokes I polarisation ($\text{SEFD}_I$) also increases at lower elevations and away from cardinal directions \citep{2021A&A...646A.143S}. The work to derive more accurate method to calculate $\text{SEFD}_I$ is on-going (Sutinjo et al. in preparation).

\section{SUMMARY AND FUTURE PLANS}
\label{sec:discussion}

\begin{figure*}

   \includegraphics[width=\textwidth]{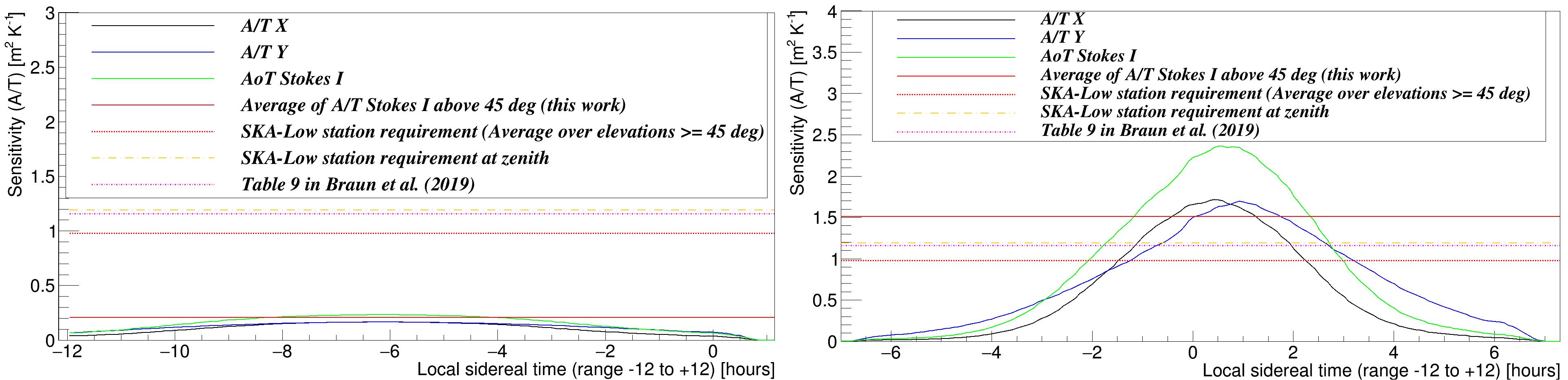}
   \caption{\red{The sensitivity of AAVS2 as a function of LST for the two extreme scenarios in terms of the sky noise. Left: for the Galactic Centre (``hot sky''), which is at an elevation $\ge$45\degree \,in the approximate LST range between $-$9.6\,h and $-$2.8\,h. Right: for the center of the Epoch of Reionisation 0 (EoR0) field (``cold sky''), which is at an elevation $\ge$45\degree\, in the approximate LST range between $-$3.3\,h and $+$3.3\,h. Solid curves show results of this work, orange and red dashed curves show SKA-Low specifications at zenith and averaged over elevations $\ge$45\degree \,respectively. The magenta dotted curve is the sensitivity from Table 9 in \citet{2019arXiv191212699B} (also averaged over elevations $\ge$45\degree). In the case of the EoR0 field, although the sensitivity averaged over elevations $\ge$45\degree\, is similar to values in the SKA-Low specification and Table 9 in \citet{2019arXiv191212699B}, it can be even up to a factor of $\sim$2.5 higher and several times lower at the highest and lowest elevations respectively (right image). Moreover, in the extreme case of the sources near the Galactic Centre (left image), the sensitivity is several times lower and never reaches the specifications and the values in \citet{2019arXiv191212699B}.}}
\label{fig_example_sources}
\end{figure*}

    
     

\begin{figure}
   \includegraphics[width=\columnwidth]{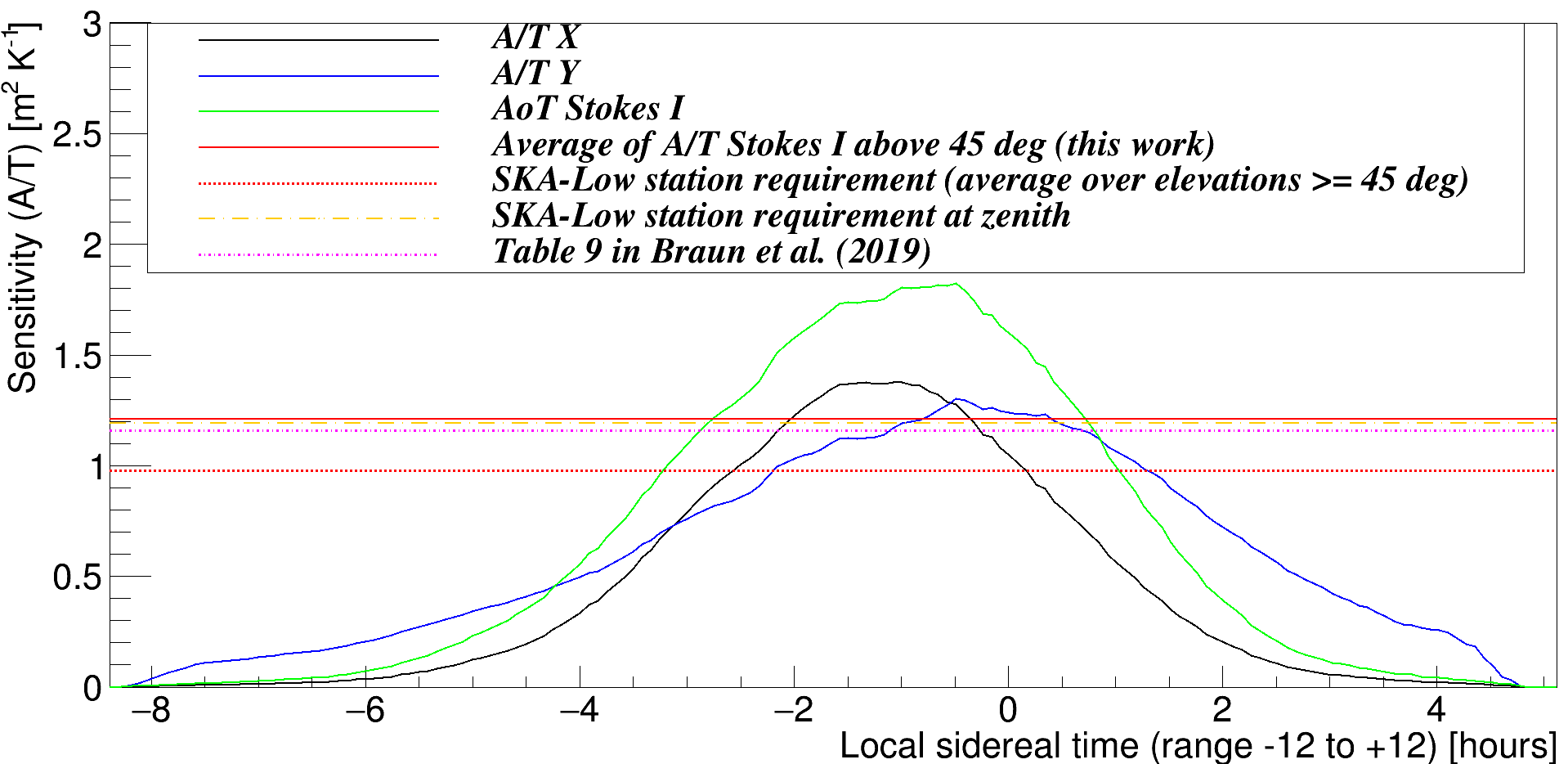}
   \caption{\red{The sensitivity of AAVS2 as a function of LST for the radio source 3C444, which is at an elevation $\ge$45\degree \,in the approximate LST range between $-$5\,h and $+$1.4\,h. Similarly to the case of the EoR0 field (Fig.~\ref{fig_example_sources}) this example shows that although the sensitivity averaged over elevations $\ge$45\degree\, is similar to values in the SKA-Low specification and Table 9 in \citet{2019arXiv191212699B}, it can be nearly a factor of 2 higher or several times lower at the best and worst observing scenarios, respectively.}}
   \label{fig_example_3c444}
\end{figure}

Accurate predictions of the station sensitivity for arbitrary pointing directions, observing times, and frequencies will be critical for planning future observations with the SKA-Low radio telescope. In order to move beyond the simplistic characterisation of sensitivity described in the current SKA-Low specifications, the complications inherent in low frequency aperture arrays that utilise closely packed, complex, and broadband antennas needs to be recognised and represented.  With this in mind, we have developed the first realistic sensitivity calculator for the future SKA-Low radio telescope. The software has already been used to verify the agreement between the expected and measured sensitivities of the two existing prototype stations of the SKA-Low (AAVS2 and EDA2), which at the same time validated the sensitivity calculator software.

Our calculator uses a database of sensitivity values calculated for two station designs (EDA2 and AAVS2), which have been operating at the Murchison Radio-astronomy Observatory since 2019. The sensitivities were pre-computed over the entire visible sky in 5\degree \, resolution, \sfrac{1}{2}\,hour intervals, and 10\,MHz steps at SKA-Low observing frequencies (50 -- 350\,MHz). The sensitivity values can be accessed both via the web interface (at \skacalcwebpage) and using a command line \textsc{python} \texttt{sensitivity\_db.py} script. The \textsc{python} package also contains a script (\texttt{eda\_sensitivity.py}), which allows the calculation of SKA-Low station sensitivity at arbitrary pointing direction, time, and frequency. The database was populated using the array factor simulation method, which is expected to be accurate at elevations above 50\degree \,\,and less accurate but useful below this elevation. More precise calculations using electromagnetic simulation software packages based on Method of Moments, such as FEKO or GALILEO are much more time consuming and so far have been completed at only a few selected frequencies. However, once full simulations at multiple frequencies are completed, the database can be updated with the resulting, more accurate, sensitivity predictions.

The package is intended to help engineers and astronomers explore approaches to SKA-Low observation planning. You can calculate the sensitivity for your favourite astronomical radio source. \red{The two extreme examples of such calculations are presented in Figure~\ref{fig_example_sources} showing the sensitivity as a function of LST calculated using our package for sources near the Galactic Centre (``hot sky'') and EoR0 field (``cold sky''). We also calculated sensitivity value averaged over elevations $\ge$45\degree\, and over-plotted with the SKA-Low specifications and predictions provided in Table 9 in \citet{2019arXiv191212699B}. In the case of the EoR0 field, although the sensitivity averaged over elevations $\ge$45\degree\, is similar to values in the SKA-Low specification and Table 9 in \citet{2019arXiv191212699B}, it can be even up to a factor of $\sim$2.5 higher and several times lower at the highest and lowest elevations respectively. Similar conclusions can be drawn for the sensitivity predictions for the 3C444 radio source (Fig.~\ref{fig_example_3c444}). Furthermore, in the extreme case of the sources near the Galactic Centre (left image), the sensitivity is several times lower and never reaches the specifications and the values in \citet{2019arXiv191212699B}.
All these examples clearly demonstrate that simple sensitivity estimates can be very inaccurate and complexities of the observations with low-frequency arrays, such as highly non-uniform sky noise, have to be taken into account in order to accurately predict sensitivity of the future observations. Hence, the answer to the question in the title is indeed ``it depends''!}


\begin{acknowledgements}
\end{acknowledgements}

This scientific work makes use of the Murchison Radio-astronomy Observatory, operated by CSIRO. We acknowledge the Wajarri Yamatji people as the traditional owners of the Observatory site. We acknowledge the Pawsey Supercomputing Centre which is supported by the Western Australian and Australian Governments.  AAVS2 and EDA2 are hosted by the MWA under an agreement via the MWA External Instruments Policy.

\vspace{0.5cm}
\begin{appendices}
\section{Examples of command line commands}
\label{appendix_cmdline_options}

Examples of typical command line options provided by the \texttt{sensitivity\_db.py} script:

\begin{itemize}
\item \textbf{Sensitivity vs. frequency} at the specified pointing direction and observation LST or UTC time. The following command line will generate two \textsc{text} files azim0\_za30\_lst15.4\_XX.txt and azim0\_za30\_lst15.4\_YY.txt with sensitivity values as a function of frequency at pointing direction (azimuth, zenith angle) = (0\degree , 30\degree) and LST = 15.4\,hours: \texttt{python ./sensitivity\_db.py -{}-azim\_deg=0 -{}-za\_deg=30 -{}-lst=15.4 -{}-out\_file="azim0\_za30\_lst15.4" -{}-do\_plot}. The optional switch \texttt{-{}-do\_plot} enables generation of \textsc{png} images. \\
      
\item \textbf{All-sky sensitivity map} at a specified observing frequency and LST or UTC time. The following command line will generate an all-sky sensitivity map (in the \textsc{png} and FITS file formats) at the time of Galactic transit: \texttt{python ./sensitivity\_db.py -{}-freq\_mhz=154.88 -{}-lst\_hours=17.76 -{}-do\_plot}. The optional parameter \texttt{-{}-save\_text} will also save the sensitivity values to the output \textsc{text} file, which maybe quite large and hence is not a default option. The name of the output file can be specified with the \texttt{-{}-out\_file} parameter. \\

\item \textbf{Sensitivity vs. time} at the specified pointing direction and observing frequency. The following command line will generate two \textsc{text} files with 24 hours of sensitivity vs. time data at zenith pointing direction, frequency 154.88\,MHz starting at 2020-02-25 02:28:49 UTC (hence unix time 1582597729) : 
\begin{enumerate}

\item Pointing direction specified in horizontal coordinates (azimuth and zenith angle) : \texttt{python ./sensitivity\_db.py -{}-freq\_mhz=154.88 -{}-unixtime\_start=1582597729  -{}-interval=86400 -{}-azim\_deg=0 -{}-za\_deg=0 -{}-do\_plot} \\

\item Pointing direction specified in equatorial coordinates (right ascension and declination) using pre-computed sensitivity values from the database : \texttt{python ./sensitivity\_db.py -{}-freq\_mhz=160.00 -{}-unixtime\_start=1636798650 -{}-interval=30 -{}-ra=333.607 -{}-dec=$-$17.026 -{}-do\_plot -{}-station=AAVS2 -{}-outfile="3C444\_sensitivity\_ux1636798650" -{}-save\_text\_file} \\

\item Pointing direction in equatorial coordinates (right ascension and declination) calculation (not using the database): \texttt{python ./eda\_sensitivity.py -{}-freq=160 -p None -g 1320833867  -m analytic -{}-ra=333.607 -{}-dec=$-$17.0266 -{}-outsens\_file="3C444\_aavs2\_sensitivity" -{}-outfile\_mode=a -{}-trcv\_type=trcv\_from\_skymodel\_with\_err  -{}-nos11 -{}-header=HEADER  -{}-use\_beam\_fits -{}-station\_name=SKALA4 -{}-size=512 -{}-trcv\_type=trcv\_aavs2\_vs\_za\_deg -{}-antenna\_locations=antenna\_locations\_aavs2.txt} \\
\end{enumerate}

Alternatively, the time range can be specified as an LST range using parameters: \texttt{-{}-lst\_start=0.00 -{}-lst\_end=24.00}.
\end{itemize}

By default the data are generated for the AAVS2 station, but this can be modified using parameter \texttt{-{}-station\_name}. The comprehensive description of the command line options can be found in documentation files included in \textsc{github} repository.

\end{appendices}

\bibliographystyle{pasa-mnras}

%
\input{skalow_sensitivity_references.bbl}

\end{document}